\documentclass[12pt,a4paper,spanish,twoside]{article}
\usepackage{amsmath, amssymb, setspace, mathrsfs} %Cosas de mates y fijar espaciado entre líneas
\usepackage{babel} %Elige el idioma de las palabras que se generan automáticamente (como tabla o Figura)
\usepackage[utf8]{inputenc} %Permite introducir directamente acentos: á en lugar de \'a etc.
\usepackage{graphicx} %Cosas chachis para Figuras
\usepackage{caption} %Cosas chachis en los nombres de las figuras
\usepackage{subcaption} %Ídem que Caption pero con subfiguras
\usepackage{float} %Permite manejar/mover/partir mejor figuras y tablas
\usepackage{fancyhdr} %Encabezado y pie de página chachi
\usepackage[usenames,dvipsnames]{xcolor} %Colrines!!
\usepackage{mathabx} %Más símbolos matemáticos
\usepackage{mathtools} %Más símbolos matemáticos
\usepackage[toc,page]{appendix} %Incluir apéndices
\usepackage{rotating} %Para girar floating objects (texto en mbox, tablas, figuras...)
\usepackage{multicol} %Para dividir la página en columnas
\usepackage{verbatim} % comentarios
\usepackage{braket} %Permite usar bra-kets para la notación de Dirac
\usepackage{pdfpages} %Para añadir páginas de pdf externos
\usepackage{cancel} %Para tachar elementos en las fórmulas
\usepackage[nottoc]{tocbibind} %Pone LOF, LOT, Apendices en la TOC
\usepackage{titlesec} %Edit title format
\usepackage{slashed} %Dirac slash notation:  \slashed{p}
\usepackage{bm} % \bm{#} does in equations what {\bf #} in text mode
\usepackage{tikz}
	\usetikzlibrary{arrows,shapes}
	\usetikzlibrary{trees}
	\usetikzlibrary{matrix,arrows} 				% For commutative diagram
	\usetikzlibrary{positioning}				% For "above of=" commands
	\usetikzlibrary{calc,through}				% For coordinates
	\usetikzlibrary{decorations.pathreplacing}  % For curly braces
\usepackage{pgffor}							% For repeating patterns
	\usetikzlibrary{decorations.pathmorphing}	% For Feynman Diagrams
	\usetikzlibrary{decorations.markings}
\tikzset{
vector/.style={decorate, decoration={snake}, draw},
provector/.style={decorate, decoration={snake,amplitude=2.5pt}, draw},
antivector/.style={decorate, decoration={snake,amplitude=-2.5pt}, draw},
fermion/.style={draw=black, postaction={decorate},
	decoration={markings,mark=at position .55 with {\arrow[draw=black]{>}}}},
fermionbar/.style={draw=black, postaction={decorate},
	decoration={markings,mark=at position .55 with {\arrow[draw=black]{<}}}},
fermionnoarrow/.style={draw=black},
gluon/.style={decorate, draw=black,
	decoration={coil,amplitude=4pt, segment length=5pt}},
scalar/.style={dashed,draw=black, postaction={decorate},
	decoration={markings,mark=at position .55 with {\arrow[draw=black]{>}}}},
scalarbar/.style={dashed,draw=black, postaction={decorate},
	decoration={markings,mark=at position .55 with {\arrow[draw=black]{<}}}},
scalarnoarrow/.style={dashed,draw=black},
electron/.style={draw=black, postaction={decorate},
	decoration={markings,mark=at position .55 with {\arrow[draw=black]{>}}}},
bigvector/.style={decorate, decoration={snake,amplitude=4pt}, draw},
}
\tikzstyle{block} = [draw, rectangle, 
    minimum height=3em, minimum width=6em]

\usepackage[pdftex, colorlinks=true,linkcolor=blue, urlcolor=blue, linktocpage]{hyperref}
\spanishdecimal{.}
\def\d{\mathrm{d}}  %para definir el operador diferencial como \d

\newcommand{\Lagr}{\mathscr{L}}

\titleformat
{\chapter} % command
[display] % shape
{\bfseries\Large} % format
{  } % label
{-1cm} % sep
{ \thechapter.  } % before-code
[ \vspace{0.5ex} ] % after-code

\newcommand{\HRule}{\rule{\linewidth}{0.5mm}}

\begin{document}
\renewcommand{\tablename}{\bf Table} %Cambia la palabra "Cuadro" por "Tabla"
\renewcommand{\figurename}{\bf Figure}
\renewcommand{\listtablename}{List of Tables} %Cambia el título "Índice de cuadros" por "Índice de Tablas
\renewcommand*\contentsname{Table of Contents} %Cambia el título "Índice General" por "Índice"
\renewcommand \listfigurename{List of Figures} %Cambia el título "Índice de figuras" por "Índice de Figuras"
\renewcommand{\appendixname}{Appendixes}
\renewcommand{\appendixtocname}{Appendixes}
\renewcommand{\appendixpagename}{Appendixes}
%\numberwithin{figure}{section}	%Estos comandos generan numeración Sección.Numero de Figura/tabla/ecuación
%\numberwithin{table}{section}
\numberwithin{equation}{section}
\renewcommand{\bibname}{References}

\def\nicefrac#1#2{\leavevmode%
    \raise.5ex\hbox{\small #1}%
    \kern-.1em/\kern-.15em%
    \lower.25ex\hbox{\small #2}}

%Estilo especial para pág. de principio de sección
\fancypagestyle{newstyle}{
\fancyhf{} % clear all header and footer fields
\fancyfoot[OC, EC]{\thepage} % except the center
\renewcommand{\headrulewidth}{0pt}
\renewcommand{\footrulewidth}{0pt}}

% Vector
\renewcommand{\vec}[1]{\bm{#1}}

% Vectror Nabla
\newcommand{\grad}{ \bm{\nabla} }

% Encabezado y pie de página
\pagestyle{fancy}
\renewcommand{\sectionmark}[1]{\markboth{#1}{}}
\renewcommand{\headrulewidth}{0pt}  %Grosor de la línea bajo el encabezado
\fancyhf{} %limpiar los campos
% página ... R=derecha, L=izquierda, C=centro; O=impar, E=par
% El número de la página es \thepage
%\fancyhead[RO]{\sectionmark}
%\fancyheadoffset{1cm}
\fancyhead[LE]{\thepage}
\fancyhead[RO]{\thepage}
%\fancyhead[CE]{\nouppercase{\thechaptr. \leftmark}}
\fancyhead[LO]{Section \thesection. \leftmark}
\fancyhead[RE]{A. Segarra}

%%%%%%%%%%%%%%%%%%%%%%%%%%%%%%%%%%%%%%%%%%%%%%
%								PORTADA
%%%%%%%%%%%%%%%%%%%%%%%%%%%%%%%%%%%%%%%%%%%%%%
\begin{titlepage}
\begin{center}
 %Upper part of the page. The '~' is needed because \\
 %only works if a paragraph has started.
%\includegraphics[width=0.15\textwidth]{./logo}~\\[1cm]
\includegraphics[width = 0.4\textwidth]{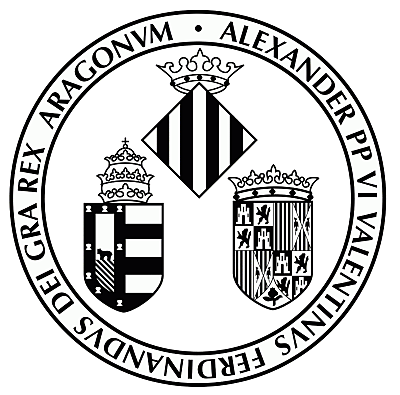}
\includegraphics[width = \textwidth]{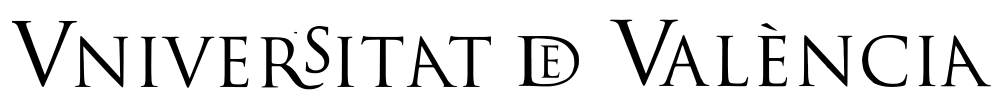}
\includegraphics[width = 0.7\textwidth]{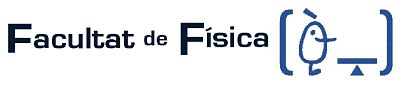}\\[1cm]
\textsc{\Large Master Thesis}\\[0.5cm]
% Title
\HRule \\[0.4cm]
{ \huge \bfseries Neutrino-Pair Exchange Long-Range Force Between Aggregate Matter \\[0.4cm] }
\HRule \\[1.5cm]
% Author and supervisor
\noindent
\begin{minipage}{0.4\textwidth}
\begin{flushleft} \large
\textsc{Author:}\\
Alejandro Segarra

\end{flushleft}
\end{minipage}%
\begin{minipage}{0.4\textwidth}
\begin{flushright} \large
\textsc{Supervisor:}\\
José Bernabéu
\end{flushright}
\end{minipage}

\vfill

July 2015

% Bottom of the page
%{\large July 2015}
\end{center}
\end{titlepage}

%\includepdf[pages={1}]{portada_TFM.pdf}

\newpage\null\thispagestyle{empty}\newpage

%\newpage
%\thispagestyle{plain}
%\begin{abstract}
%    As most research papers have an abstract, there are predefined
%    commands for telling LaTeX which part of the content makes up the
%    abstract. This should appear in its logical order, therefore,
%    after the top matter, but before the main sections of the
%    body. This command is available for the document classes article
%    and report, but not book.
%\end{abstract}
%
%
%\newpage\null\thispagestyle{empty}\newpage

\newpage
\thispagestyle{newstyle}
\addcontentsline{toc}{section}{\hspace{0.6cm}Abstract}
\section*{Abstract}

We study the long-range force arising between two neutral---of
electric charge---aggregates of matter due to a neutrino-pair
exchange, in the limit of zero neutrino mass. The conceptual basis for
the construction of the effective potential comes from the coherent
scattering amplitude at low values of $t$. This amplitude is obtained
using the methodology of an unsubtracted dispersion relation in $t$ at
threshold for $s$, where $(s,\, t)$ are the Lorentz invariant
scattering variables. The ultraviolet behavior is irrelevant for the
long-range force. In turn, the absorptive part in the $t$-dependence
is given by the corresponding unitarity relation. We show that the
potential describing this force decreases as $r^{-5}$ at large
separation distance $r$. This interaction is described in terms of its
own charge, which we call the weak flavor charge of the interacting
systems, that depends on the flavor of the neutrino as $Q_W^e = 2Z-N$,
$Q_W^\mu = Q_W^\tau = -N$. The flavor dependence of the potential
factorizes %, in the massless neutrino limit,
in the product of the weak charges of the interacting systems, so that
the resulting force is always repulsive. Furthermore, this charge is
proportional to the number of constituent particles, which differs
from the global mass, so this interaction could be disentangled from
gravitation through deviations from the Equivalence Principle.

\newpage\null\thispagestyle{empty}\newpage

\newpage
\thispagestyle{newstyle}
\tableofcontents
%\addtocontents{toc}{~\hfill\textbf{\underline{Page}}\par}

%%%%%%%%%%%%%%%%%%%%%%%%%%%%%%%%%%%%%%%%%%%%%%
%                     MAIN
%%%%%%%%%%%%%%%%%%%%%%%%%%%%%%%%%%%%%%%%%%%%%%
\newpage
\thispagestyle{newstyle}
\section{Introduction}

It's been 85 years since Wolfgang Pauli postulated the existence of
the neutrino in order to explain the continuous spectrum in
$\beta$-decays, and 59 years since Reines and Cowan discovered it. In
those years, we've learnt many properties about this particle, such as
the fact that it only interacts through weak interactions---all of its
charges but weak isospin are zero. In fact, in the framework of the
Standard Model \cite{pichEW}, there are only left-handed neutrinos, so
Standard Model neutrinos are massless---we can't generate a neutrino
mass through a Yukawa-type coupling with a Higgs doublet.

Other interesting phenomena related to this particle are neutrino
oscillations \cite{taup}, which have been well established
experimentally since 1998. This process is understood as the fact that
there is a mismatch between mass eigenstates and flavor eigenstates,
so that flavors get mixed along free propagation.  Indeed, the
observation of neutrino oscillations is a direct measurement of the
mass difference between the three states, proving that neutrinos are
massive particles, which is a first signal of Physics beyond the
Standard Model.

Therefore, the study of the origin of neutrino mass is one of the
directions in which we can expect finding new Physics, even though its
small value ($m_\nu \lesssim 1$ eV \cite{PDG}) makes it hard to
observe experimentally. As well as determining the absolute mass of
the neutrino, there's still a more fundamental question about their
nature unanswered: since neutrinos can be neutral of all charges,
their finite mass could be explained through a Dirac mass term
(implying that neutrinos and antineutrinos are different particles,
described by $4-$component Dirac spinors) or through a Majorana one
(implying that neutrinos are self-conjugate of all charges, described
by $2$ independent degrees of freedom).

In any case, the fact that their masses are very low stands, and we
discuss here another property of neutrinos as mediators of a new
force. As is well known, the processes represented in Quantum Field
Theory by the exchange of a massless particle give raise to long-range
interactions. An easy example is the scattering of two particles
mediated by a photon, which---at tree level---describes Coulomb
scattering. Our objective in this work is the application of these
ideas to a process mediated by neutrinos. According to the Electroweak
Lagrangian, the lowest-order process is a neutrino-pair exchange,
which---since neutrinos are nearly massless---describes an interaction
of long range.

With this idea in mind, we review in Section \ref{sec:Potentials} the
relation between the Feynman amplitude in Born approximation and an
effective potential, which is a Fourier Transform. The amplitude at
low $t$, associated to the long-range behavior, is obtained by means
of an unsubtracted dispersion relation. Its ultraviolet dependence is
of no relevance. In order to simplify the calculation of the
potential, in Section \ref{sec:unitarity} we exploit the untitarity of
the $S$ matrix, writing the absorptive part of the $1-$loop scattering
amplitude with the amplitude of the tree-level scattering process.

In Section \ref{sec:Leff}, we study the low-energy limit of the
Electroweak Lagrangian in terms of a contact interaction, establishing
the framework for the calculation of the scattering amplitude
including both neutral current and charged current vertices. We
compute in detail this amplitude in Section \ref{sec:amplitude}, where
it's natural to introduce the concept of a weak flavor charge of
matter. In terms of this amplitude, obtaining the interaction
potential is straightforward, and we find in Section \ref{sec:weakpot}
that it leads to a repulsive force which decreases as $r^{-6}$.

We conclude this work analyzing the possibility of an experimental
measurement of this interaction, which is relevant between nanometers
and microns, where there are also residual electromagnetic
interactions---such as Van der Waals or Casimir-Polder forces--- and
gravitation. The measurement of this weak interaction is very
compelling, since it could give information about properties of the
neutrino such as its absolute mass, which is still unknown, or it
could even help us to answer the most fundamental question regarding
neutrinos, whether they are Dirac or Majorana particles. These points
are considered in Sections \ref{sec:conclusions} and
\ref{sec:prospects}.

\newpage
\thispagestyle{newstyle}
\section{From a Quantum Field Theory to an Effective Potential}
\label{sec:Potentials}
We are interested in calculating the interaction potential resulting
from a neutrino-pair exchange between aggregates of matter, which is
an interaction described in the framework of a Quantum Field
Theory. Therefore, we will begin this work relating the concepts of
interaction potential and Feynman amplitude.

\subsection{The Coulomb potential}
\label{sec:Coulomb}

It is known that the interaction between two electrically charged
particles, say $A$ and $B$, is described by the Coulomb potential,
\begin{equation}
	V_C(r) = \frac{e^2}{4\pi}\, \frac{Q_A Q_B}{r}\,,
\end{equation}
where $e$ is the charge of the proton, $Q_J$ the charge of the
particle $J$ in units of $e$ and $r$ the distance between the two
particles. Throughout this work, we'll use the Natural System of Units
and the Heaviside electric system---all conventions are stated in
Appendix \ref{sec:appA}.

We are interested in calculating this potential using the Quantum
Electrodynamics (QED), which is described by the interaction
Lagrangian
\begin{equation}
	\Lagr_\text{QED} = -eQ\, \bar \psi\gamma^\mu \psi\, A_\mu\,.
\end{equation}

\begin{figure}[t]
	\centering
	\begin{subfigure}{0.45\textwidth}
	\begin{tikzpicture}[line width=1.5 pt, scale=1.65]
		% A:
		\draw[fermion] (150:1.5) -- (0, 0);
			\node at (150:1.5)[above left]{$A(p_1)$};
		\draw[fermionbar] (30:1.5) -- (0, 0);
			\node at (30:1.5)[above right]{$A(p_3)$};
		\node at (0,0.05)[above]{$Q_A$};
		%Propagator
		\draw[vector] (0,0) -- (0,-1);
			\node at (0, -0.5)[right]{$\gamma (q)$};
		% B:
		\begin{scope}[shift={(0, -1)}]
			\draw[fermion] (-150:1.5) -- (0,0);
				\node at (-150:1.5)[below left]{$B(p_2)$};
			\draw[fermionbar] (-30:1.5) -- (0,0);
				\node at (-30:1.5)[below right]{$B(p_4)$};
			\node at (0, -0.05)[below]{$Q_B$};
		\end{scope}
	 \end{tikzpicture}
	 \caption{}
	 \label{fig:Coulomb}
	\end{subfigure}
	\hfill
	\begin{subfigure}{0.45\textwidth}
	\begin{tikzpicture}[line width=1.5 pt, scale=1.65]
		% A:
		\draw[fermion] (150:1.5) -- (0, 0);
			\node at (150:1.5)[above left]{$A(p_1)$};
		\draw[fermionbar] (30:1.5) -- (0, 0);
			\node at (30:1.5)[above right]{$A(p_3)$};
		%Propagator
		\draw[scalarnoarrow] (0,0) -- (0,-1);
			\node at (0, -0.5)[right]{$\phi (q)$};
		% B:
		\begin{scope}[shift={(0, -1)}]
			\draw[fermion] (-150:1.5) -- (0,0);
				\node at (-150:1.5)[below left]{$B(p_2)$};
			\draw[fermionbar] (-30:1.5) -- (0,0);
				\node at (-30:1.5)[below right]{$B(p_4)$};
		\end{scope}
	 \end{tikzpicture}
	 \caption{}
 	 \label{fig:Yukawa}
	\end{subfigure}
        \caption{Lowest-order Feynman diagrams for $AB\to AB$ elastic
          scattering. (a)~QED interaction, mediated by a photon, where
          $A$ and $B$ are particles of electric charge $Q_A$ and
          $Q_B$. (b)~Yukawa interaction, mediated by a scalar $\phi$
          of mass $\mu$.}
\end{figure}
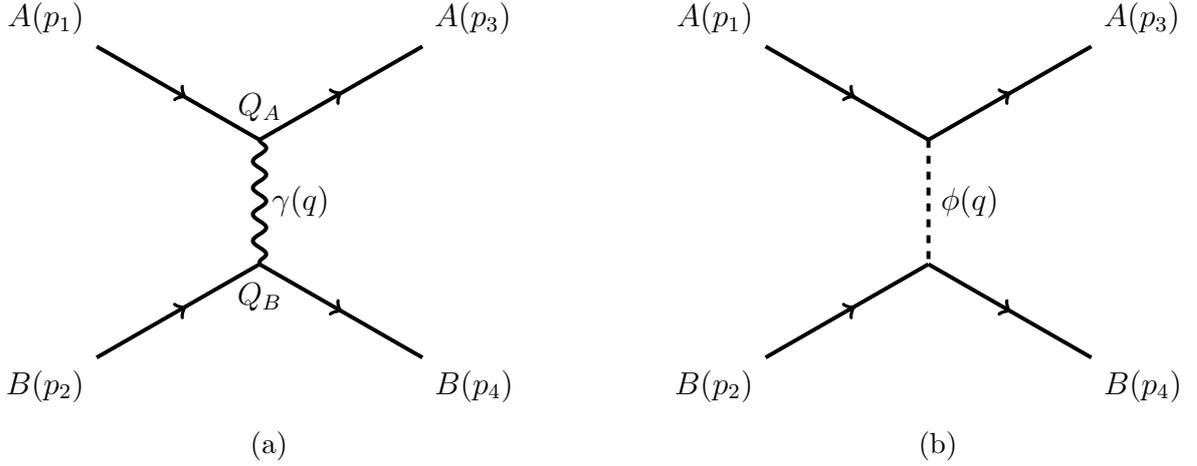

In this framework, the $AB\to AB$ elastic scattering is described---at
leading order---by the Feynman graph from Fig.\ref{fig:Coulomb}. Using
the QED Feynman rules \cite{pichEW}, the amplitude of the process is
\begin{equation}
	{\cal M} = e^2 Q_A Q_B \left[ \bar u(p_3) \gamma^\mu u(p_1) \right]\, \frac{1}{q^2} \,\left[ \bar u(p_4) \gamma_\mu u(p_2) \right]\,.
\end{equation}

Since we are looking for a long-range coherent interaction, we can
simplify
\begin{equation}
	{\cal M} \approx e^2 Q_A Q_B \left[ \bar u(p_3) \gamma^0 u(p_1) \right]\, \frac{1}{q^2} \,\left[ \bar u(p_4) \gamma_0 u(p_2) \right]\,
\end{equation}
taking into account the fact that $\gamma^0$ is related to the
electric charge, which is coherent, while $\vec{\gamma}$ is related to
the electromagnetic current, which is not a coherent quantity.

Using $\left( \gamma^0 \right)^2 = 1$ and dropping external-line
factors, we get
\begin{equation}
	M(q^2) = e^2 Q_A Q_B \, \frac{1}{q^2}\,,
\end{equation}
where we defined $M(q^2)$ as
${\cal M}(q^2) \equiv \bar u^{(A)} (p_3) \bar u^{(B)} (p_4) M(q^2)
u^{(B)} (p_2) u^{(A)} (p_1)$.

Since this is a scattering process, we can work in the Breit reference
frame (defined by $q^0 = 0$), which describes the non-relativistic
limit (low energy transfer), where
\begin{equation}
	M(q^2) = - e^2 Q_A Q_B \, \frac{1}{{\vec{q}\,}^2}\,.
\end{equation}

We can compute the 3-dimensional Fourier Transform of this quantity
(see Appendix \ref{sec:FTCoulomb}), and we find
\begin{equation}
	{\cal F}\left\{ M\right\} (r) \equiv \int \frac{\d^3 q}{(2\pi)^3}\, e^{i\vec q\, \vec r}\, M(q^2) = - \frac{e^2}{4\pi}\, \frac{Q_A Q_B}{r} = - V_C(r)\,.
\end{equation}

This expression shows the relation between the Quantum Field Theory
Feynman amplitude and the interaction potential used in a potential
description of the system dynamics. Before considering a more general
case, let's look at another simple one: the Yukawa interaction.

\newpage
\subsection{The Yukawa interaction}

Another well-known potential is Yukawa's, which describes an effective
central strong nuclear force acting between nucleons,
\begin{equation}
	V_Y(r) = - \frac{g^2}{4\pi} \frac{e^{-\mu r}}{r}\,.
\end{equation}

From a Quantum Field Theory point of view, this interaction is
described by the Lagrangian
\begin{equation}
	\Lagr_Y = -g \phi \bar \psi \psi\,,
\end{equation}
where $\phi$ is a scalar field and $\psi$ is a fermionic field. Such a
scalar can be physically associated to the $\sigma$ meson for the
interacting $\pi$-$\pi$ mediation. The $AB \to AB$ scattering
amplitude described by this Lagrangian is the one represented in
Fig.\ref{fig:Yukawa}, so it is
\begin{equation}
	{\cal M} = -g^2 \left[ \bar u(p_3) u(p_1) \right] \frac{1}{q^2-\mu^2} \left[ \bar u(p_4) u(p_2) \right]\,,
\end{equation}
where $\mu$ is mass of the scalar, and
\begin{equation}
	M(q^2) = \frac{-g^2}{q^2-\mu^2}\,.
\end{equation}

Again, we can work in the Breit reference frame, so that
\begin{equation}
	M(q^2) = \frac{g^2}{\vec {q\,}^2 + \mu^2}\,.
\end{equation}

The potential must be related to the Fourier Transform of this
$M(q^2)$, which is also calculated in Appendix \ref{sec:FTCoulomb},
\begin{equation}
	{\cal F}\left\{ M\right\} (r) = \frac{g^2}{4\pi} \frac{e^{-\mu r}}{r} = - V_Y(r)\,,
\end{equation}
which is the same relation between $M(q^2)$ and $V(r)$ that we
obtained in the Coulomb case.

\newpage
\subsection{A more general case: particle-pair exchange}
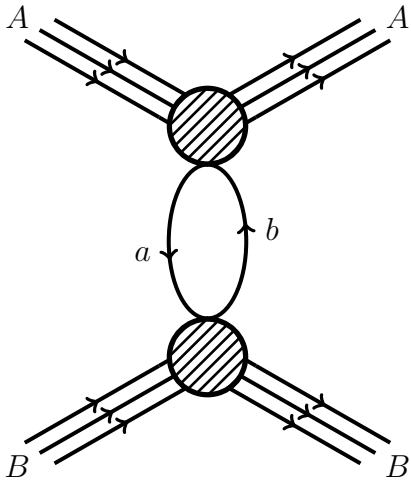
\begin{figure}[t]
    \centering
    \begin{tikzpicture}[line width=1.5 pt, scale=1.7]
        % A in:
        \draw[fermion](150:1.5) -- (150:0.3cm); \node at (150:1.7)
        {$A$};
        \begin{scope}[shift={(35:4pt)}]
            \draw[fermion](150:1.5) -- (150:0.3cm);
        \end{scope}
        \begin{scope}[shift={(35:-4pt)}]
            \draw[fermion](150:1.5) -- (150:0.3cm); \draw (150:0.3) --
            (150:0);
        \end{scope}
		
        % A out:
        \draw[fermionbar](30:1.5) -- (30:0.3cm); \node at (30:1.7)
        {$A$};
        \begin{scope}[shift={(-35:-4pt)}]
            \draw[fermionbar](30:1.5) -- (30:0.3cm);
        \end{scope}
        \begin{scope}[shift={(-35:4pt)}]
            \draw[fermionbar](30:1.5) -- (30:0.3cm); \draw (30:0.3) --
            (30:0);
        \end{scope}
		
        % A blob:
        \draw[fill=black] (0,0) circle (.3cm); \draw[fill=white] (0,0)
        circle (.29cm);
        \begin{scope}
            \clip (0,0) circle (.3cm); \foreach \x in {-.9,-.8,...,.3}
            \draw[line width=1 pt] (\x,-.3) -- (\x+.6,.3);
        \end{scope}
	  	
        % Neutrino loop
        \begin{scope}
            \clip (-2,-2) rectangle (0,0); \draw[fermion] (0,-0.9)
            ellipse (0.3 and 0.6); \node at (-0.5, -1) {$a$};
        \end{scope}
        \begin{scope}
            \clip (2,-2) rectangle (0,0); \draw[fermion] (0,-0.9)
            ellipse (-0.3 and -0.6); \node at (0.5, -0.8) {$b$};
        \end{scope}
	  				
        \begin{scope}[shift={(0,-1.8)}]
            % B in:
            \draw[fermion](-150:1.5) -- (-150:0.3cm); \node at
            (-150:1.7) {$B$};
            \begin{scope}[shift={(-35:4pt)}]
                \draw[fermion](-150:1.5) -- (-150:0.3cm);
            \end{scope}
            \begin{scope}[shift={(-35:-4pt)}]
                \draw[fermion](-150:1.5) -- (-150:0.3cm); \draw
                (-150:0.3) -- (-150:0);
            \end{scope}
			
            % B out:
            \draw[fermionbar](-30:1.5) -- (-30:0.3cm); \node at
            (-30:1.7) {$B$};
            \begin{scope}[shift={(35:-4pt)}]
                \draw[fermionbar](-30:1.5) -- (-30:0.3cm);
            \end{scope}
            \begin{scope}[shift={(35:4pt)}]
                \draw[fermionbar](-30:1.5) -- (-30:0.3cm); \draw
                (-30:0.3) -- (-30:0);
            \end{scope}
			
            % B blob:
            \draw[fill=black] (0,0) circle (.3cm); \draw[fill=white]
            (0,0) circle (.29cm);
            \begin{scope}
                \clip (0,0) circle (.3cm); \foreach \x in
                {-.9,-.8,...,.3} \draw[line width=1 pt] (\x,-.3) --
                (\x+.6,.3);
            \end{scope}
        \end{scope}
    \end{tikzpicture}
    \caption{Feynman diagram for $AB\to AB$ elastic scattering
      mediated by $a, b$ exchange.}
    \label{fig:ABAB}
\end{figure}

As we have just seen, the interaction potential between particles $A$
and $B$ is the Fourier Transform
\begin{equation}
    V (r) = - \int \frac{\d^3 q}{(2\pi)^3} \, e^{i \vec{q}\, \vec{r}} \, M(q^2)\,,
    \label{eq:defV}
\end{equation}
where $M$ is the lowest order Feynman amplitude for the process
$AB\to AB$, with both $A$ and $B$ on-shell, but without external-leg
factors, as is discussed in \cite{sucher65}\footnote{Beware a minus
  sign between their convention for the Feynman amplitude and
  ours.}. In the case of a pair exchange, this process will be the one
represented in Fig.\ref{fig:ABAB}.

In order to compute integral (\ref{eq:defV}), we rewrite the amplitude
as a dispersion relation following the steps mentioned in
\cite{rocha}. We can extend $t$ to the complex plane and expand the
amplitude using Cauchy's Formula \cite{complexa},
\begin{equation}
    f(z) = \frac{1}{2\pi i} \int_C \d z^\prime \, \frac{f (z^\prime)}{z^\prime - z}\,,
\end{equation}
which is valid whenever $f(z)$ is analytic inside $C$.

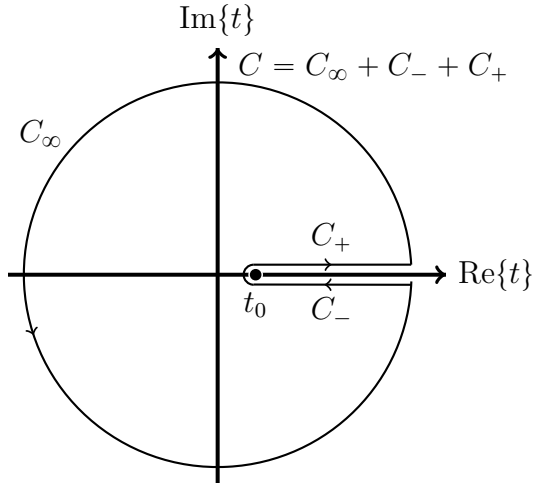
\begin{figure}[t]
    \centering
    \begin{tikzpicture}[line width = 1.5pt, scale = 1]
        % Axes:
        \begin{scope}
            \clip (-2.75,-2.75) rectangle (3,3); \draw[fermion](-3, 0)
            -- (7.91 ,0); \draw[fermion](0, -3) -- (0, 7.91);
        \end{scope}
        \node at (3, 0) [right] {Re$\{t\}$}; \node at (0,3) [above]
        {Im$\{t\}$};
        % Branching point:
        \begin{scope}[shift={(0.5,0)}]
            \draw[white, fill = white] (0,0) circle (0.07); \draw[fill
            = black] (0,0) circle (0.05); \node at
            (0,-0.075)[below]{$t_0$};
        \end{scope}
        % Integration path:
        \draw[fermion, thick] (3:2.55) arc (3:358:2.55); \node at
        (135:2.65)[left]{$C_\infty$}; \draw[fermionbar, thick]
        (3:2.55) -- (0.475, 0.133); \node at (1.5,
        0.133)[above]{$C_+$}; \draw[fermion, thick] (-3:2.55) --
        (0.475, -0.13); \node at (1.5, -0.133)[below]{$C_-$};
        \begin{scope}[shift={(0.475,0)}]
            \draw[thick] (0, 0.133) arc (90:270:0.133);
        \end{scope}
        \node at (4,3.1)[below left]{$C = C_\infty + C_- + C_+$};
    \end{tikzpicture}
    \caption{Integration path (in the complex plane of the $t$
      Mandelstam variable) used in the dispersion relation
      decomposition of the Feynman amplitude of the process, as
      discussed in the text.}
    \label{fig:C}
\end{figure}

The physical region of the $t$ variable of elastic scattering
processes has $t<0$, so we want the $\mathbb{R}^-$ axis inside
$C$. Also, the $t-$channel amplitude will have a branching point at
$t = (m_a + m_b)^2 \equiv t_0 \geq 0$, so we can use Cauchy's Formula
with the integration path shown in Fig.\ref{fig:C}. In fact, the
physical region is $-s\leq t \leq t_0$, but we are only interested in
the long-range interaction, which is associated to low values of
$\left| t \right|$. Since $\left| t \right| \sim s \sim (M_A+M_B)^2$
describes interactions of much shorter range than the nuclear size
whenever $A$ and $B$ are aggregates of matter, we can
take $s\to \infty$ without affecting the
long-range amplitude, as we have done in considering the path in
Fig.\ref{fig:C}.

If the amplitude vanishes along the $C_\infty$ circumference, as
$\left| t \right| \to \infty$, the only contribution is the one coming
from the integral on both sides of the cut along the real $t$ axis,
\begin{align}
  \nonumber M(t) &=  \frac{1}{2\pi i}\lim_{\epsilon\to 0}\int^{t_0}_{\infty} \d t^\prime\, \frac{M(t^\prime - i\epsilon)}{t^\prime- t} +\frac{1}{2\pi i} \lim_{\epsilon\to 0}\int_{t_0}^\infty \d t^\prime\, \frac{M(t^\prime + i\epsilon)}{t^\prime- t} = \\
                 & =  \frac{1}{2\pi i}\lim_{\epsilon\to 0}\int_{t_0}^\infty \d t^\prime\, \frac{M(t^\prime + i\epsilon) - M(t-i\epsilon)}{t^\prime- t}\,. \label{eq:MCinfty}
\end{align}

If not vanishing at $C_\infty$, we'd have to either rewrite the
dispersion relation for the subtracted amplitude or include the
contribution of $C_\infty$. We continue with the formulation without
subtractions, because the contribution along $C_\infty$ is of short
range. We then understand Eq.(\ref{eq:MCinfty}) for the long-range
amplitude.

In order to compute the analytically extended amplitude both above and
below the unitarity cut, we can relate them using Schwarz Reflexion
Principle \cite{complexa},
\begin{equation}
    M(t -i\epsilon) = M^*(t+i\epsilon)\,.
\end{equation}

Using this relation, we can easily write
\begin{equation}
    M(t) = \frac{1}{\pi}\int_{t_0}^\infty \, \d t^\prime\, \frac{\text{Im}\left\{M(t^\prime)\right\}}{t^\prime - t}\,,
\end{equation}
which is the so-called $t-$channel dispersion relation of the Feynman
amplitude. Putting this expression into (\ref{eq:defV}) and rewriting
$(t^\prime - t)^{-1}$ as (\ref{eq:FTSW}) states, we get
\begin{align}
  \nonumber V(r) &= \frac{-1}{4\pi^2} \int \frac{\d^3 q}{(2\pi)^3} \, e^{i \vec{q}\, \vec{r}} \, \int_{t_0}^\infty \, \d t^\prime\, \text{Im}\left\{M(t^\prime)\right\} \int \d^3r^\prime\, e^{-i \vec{q}\, {\vec{r}\,}^\prime}\, \frac{e^{-\sqrt{t^\prime} r^\prime}}{r^\prime} = \\
  \nonumber &= \frac{-1}{4\pi^2}\, \int_{t_0}^\infty \, \d t^\prime\, \text{Im}\left\{M(t^\prime)\right\} \int \d^3r^\prime\,  \frac{e^{-\sqrt{t^\prime}r^\prime}}{r^\prime} \, \delta^{(3)}(\vec r - {\vec r\,}^\prime) = \\
                 &= \frac{-1}{4\pi^2}\, \int_{t_0}^\infty \, \d t^\prime\, \text{Im}\left\{M(t^\prime)\right\}\, \frac{ e^{-\sqrt{t^\prime}r}}{r} \,.
\end{align}

Therefore, the non-relativistic potential
\begin{equation}
    \boxed{\hspace{0.25cm} V(r) = \frac{-1}{4\pi^2 r}\, \int_{t_0}^\infty \, \d t^\prime\, \text{Im}\left\{M(t^\prime)\right\}\,  e^{-\sqrt{t^\prime}r} \hspace{0.25cm} }
    \label{eq:VIm}
\end{equation}
is determined by the absorptive part of the Feynman amplitude. Since
we are not interested in the whole $M(t)$, but only in the
Im$\{ M(t)\}$, we can make a profit from the unitarity of the $S$
matrix to simplify our calculations.

\newpage
\thispagestyle{newstyle}
\section{Unitarity Relation. Absorptive Part}
\label{sec:unitarity}

Physical processes are determined by matrix elements of the scattering
matrix $S$. The $S$ matrix relates the orthonormal basis of initial
states with the final states' one, so it has to be a unitary operator,
\begin{equation}
	S^\dagger S = 1.
	\label{eq:1}
\end{equation}

We define the reduced scattering matrix $T$ as $S \equiv 1+i\,T$,
which describes processes where there really is an
interaction---initial and final states are not the same ones. In terms
of this operator, the unitarity relation (\ref{eq:1}) becomes
\[
	1 = S^\dagger S = (1-i\,T^\dagger)(1+i\,T) = 1-i\,T^\dagger +i\,T+T^\dagger T,
\]
so
\begin{equation}
	-i(T-T^\dagger) = T^\dagger T\,.
\end{equation}

In order to describe a physical process, we have to consider the
matrix element
$\bra{f}S-1\ket{i} = i\, \bra{f} T \ket{i} \equiv i\, (2\pi)^4
\delta^{(4)}(p_f-p_i) {\cal M}(i\to f)$,
where $\ket{i}$ is the initial state and $\ket{f}$ is the final
one. Therefore, we need to sandwich the previous relation between
those states---we begin computing the left-hand side (LHS),
\begin{align}
	\nonumber \bra{f} \text{LHS} \ket{i} &= -i\,\bra{f} T-T^\dagger \ket{i} =\\
	\nonumber	&= -i \left[ \bra{f} T \ket{i} - \bra{i} T \ket{f}^* \right] = \\
		&= -i\, \times 2i \,\text{Im} \left\{  \bra{f} T \ket{i} \right\}\,,
\end{align}
where we assumed that time reversal is a good symmetry to write
\begin{equation}
	T(i\to f) - T(f\to i)^* = 2\,\text{Im}\left\{ T(i\to f) \right\}\,.
\end{equation}

On the other hand,
\begin{align}
	\nonumber \bra{f} \text{RHS} \ket{i} &= \bra{f} T^\dagger T \ket{i} =\\
	\nonumber	&= \bra{f} T^\dagger \left[ \sum_n \int \prod_{j=1}^n \frac{\d^3 q_j}{(2\pi)^3 2E_{q_j}} \ket{q_n}\bra{ q_n} \right] T \ket{i} =\\
		&= \sum_n \int \prod_{j=1}^n \frac{\d^3 q_j}{(2\pi)^3 2E_{q_j}} \bra{f} T^\dagger \ket{ q_n}\bra{ q_n}  T \ket{i}\,,
		%&= (2\pi)^8 \,\delta^{(4)}(p_f-\sum_{j=1}^nq_j )\, \delta^{(4)}(p_i-\sum_{j=1}^nq_j)\, \sum_n \int \prod_{j=1}^n \frac{\d^3 q_j}{(2\pi)^3 2E_{q_j}} {\cal M}(i\to  q_n) {\cal M}^*(f\to  q_n)\,,
\end{align}
where in the second line we have inserted an identity---a sum over all
possible states, with $\ket{q_n}$ representing a state of $n$
particles with 4-momenta $q_1, q_2... \,q_n$.

Now we can write the unitarity relation
$\bra{f} \text{LHS} \ket{i} = \bra{f} \text{RHS} \ket{i}$ as
\begin{equation}
	\boxed{ \hspace{0.25cm}\begin{aligned} \mbox{}\\ \mbox{} \end{aligned} \text{Im} \left\{  \bra{f} T \ket{i} \right\} = \frac{1}{2} \sum_n \int \d Q_n \bra{q_n} T \ket{f}^*\bra{ q_n}  T  \ket{i} \hspace{0.25cm} }\,.
	\label{eq:unit}
\end{equation}
%\begin{equation}
%	2\,\text{Im} \left\{ {\cal M}(i\to f) \right\} =\sum_n \int \prod_{j=1}^n \frac{\d^3 q_j}{(2\pi)^3 2E_{q_j}} \, (2\pi)^4 \,\delta^{(4)}(p_f-\sum_{l=1}^nq_l)\,{\cal M}(i\to \vec{q}_n) {\cal M}^*(f\to \vec{q}_n)\,.
%\end{equation}

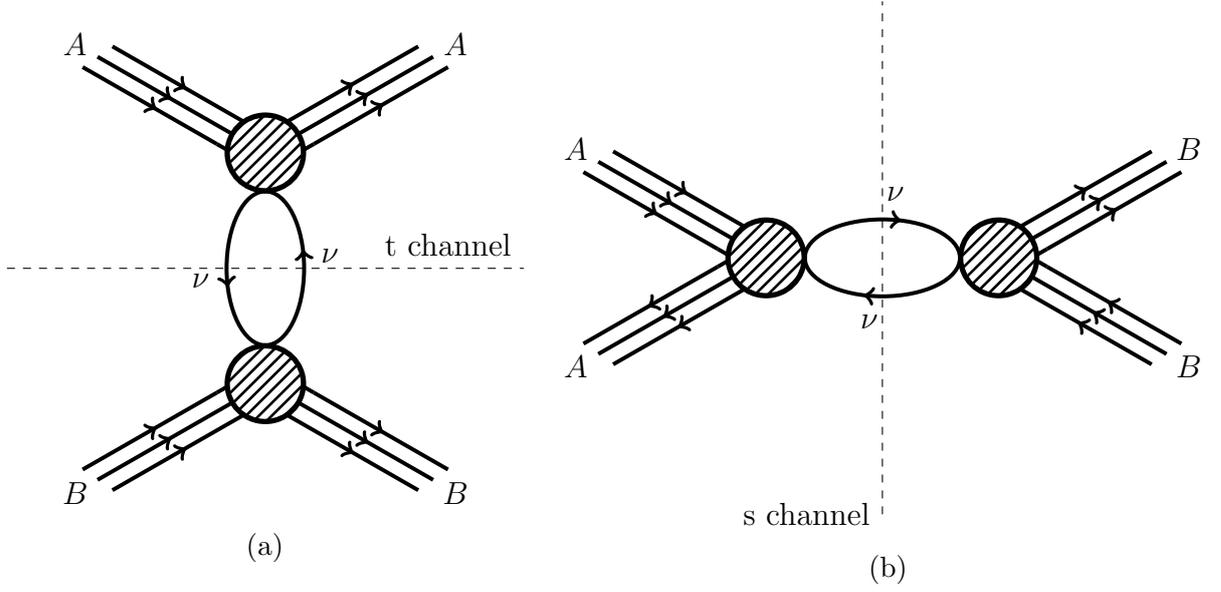
\begin{figure}[t]
	\centering
	\begin{subfigure}{0.425\textwidth}
	\centering
	\begin{tikzpicture}[line width=1.5 pt, scale=1.7]
		% A in:
		\draw[fermion](150:1.5) -- (150:0.3cm);
			\node at (150:1.7) {$A$};
		\begin{scope}[shift={(35:4pt)}]
			\draw[fermion](150:1.5) -- (150:0.3cm);
		\end{scope}
		\begin{scope}[shift={(35:-4pt)}]
			\draw[fermion](150:1.5) -- (150:0.3cm);
			\draw (150:0.3) -- (150:0);
		\end{scope}
		
		% A out:
		\draw[fermionbar](30:1.5) -- (30:0.3cm);
			\node at (30:1.7) {$A$};
		\begin{scope}[shift={(-35:-4pt)}]
			\draw[fermionbar](30:1.5) -- (30:0.3cm);
		\end{scope}
		\begin{scope}[shift={(-35:4pt)}]
			\draw[fermionbar](30:1.5) -- (30:0.3cm);
			\draw (30:0.3) -- (30:0);
		\end{scope}
		
		% A blob:
		\draw[fill=black] (0,0) circle (.3cm);
		\draw[fill=white] (0,0) circle (.29cm);
		\begin{scope}
		    	\clip (0,0) circle (.3cm);
		    	\foreach \x in {-.9,-.8,...,.3}
				\draw[line width=1 pt] (\x,-.3) -- (\x+.6,.3);
	  	\end{scope}
	  	
	  	% Neutrino loop
	  	\begin{scope}
	  		\clip (-2,-2) rectangle (0,0);
		  	\draw[fermion] (0,-0.9) ellipse (0.3 and 0.6);
		  	\node at (-0.5, -1) {$\nu$};
	  	\end{scope}
	  	\begin{scope}
	  		\clip (2,-2) rectangle (0,0);
		  	\draw[fermion] (0,-0.9) ellipse (-0.3 and -0.6);
		  	\node at (0.5, -0.8) {$\nu$};
	  	\end{scope}
	  	
	  	%Unitarity cut:
	  	\draw[dashed, thin] (-2,-0.9) -- (2, -0.9);
	  	\node at (2, -0.9) [above left] {t channel};
			
	  	\begin{scope}[shift={(0,-1.8)}]
		  	% B in:
			\draw[fermion](-150:1.5) -- (-150:0.3cm);
				\node at (-150:1.7) {$B$};
			\begin{scope}[shift={(-35:4pt)}]
				\draw[fermion](-150:1.5) -- (-150:0.3cm);
			\end{scope}
			\begin{scope}[shift={(-35:-4pt)}]
				\draw[fermion](-150:1.5) -- (-150:0.3cm);
				\draw (-150:0.3) -- (-150:0);
			\end{scope}
			
			% B out:
			\draw[fermionbar](-30:1.5) -- (-30:0.3cm);
				\node at (-30:1.7) {$B$};
			\begin{scope}[shift={(35:-4pt)}]
				\draw[fermionbar](-30:1.5) -- (-30:0.3cm);
			\end{scope}
			\begin{scope}[shift={(35:4pt)}]
				\draw[fermionbar](-30:1.5) -- (-30:0.3cm);
				\draw (-30:0.3) -- (-30:0);
			\end{scope}
			
	  		% B blob:
			\draw[fill=black] (0,0) circle (.3cm);
			\draw[fill=white] (0,0) circle (.29cm);
			\begin{scope}
			    	\clip (0,0) circle (.3cm);
			    	\foreach \x in {-.9,-.8,...,.3}
					\draw[line width=1 pt] (\x,-.3) -- (\x+.6,.3);
		  	\end{scope}
	  	\end{scope}
	 \end{tikzpicture}
	 \caption{}
	 \label{fig:ABABt}
	 \end{subfigure}
	 \hfill
	\begin{subfigure}{0.55\textwidth}
	\begin{tikzpicture}[line width=1.5 pt, scale=1.7]
		% A in:
		\draw[fermion](150:1.5) -- (150:0.3cm);
			\node at (150:1.7) {$A$};
		\begin{scope}[shift={(35:4pt)}]
			\draw[fermion](150:1.5) -- (150:0.3cm);
		\end{scope}
		\begin{scope}[shift={(35:-4pt)}]
			\draw[fermion](150:1.5) -- (150:0.3cm);
			\draw (150:0.3) -- (150:0);
		\end{scope}
		
		%\bar A in:
		\draw[fermionbar](210:1.5) -- (210:0.3cm);
			\node at (210:1.7) {$A$};
		\begin{scope}[shift={(-35:4pt)}]
			\draw[fermionbar](210:1.5) -- (210:0.3cm);
		\end{scope}
		\begin{scope}[shift={(-35:-4pt)}]
			\draw[fermionbar](210:1.5) -- (210:0.3cm);
			\draw (210:0.3) -- (210:0);
		\end{scope}
		
		% A blob:
		\draw[fill=black] (0,0) circle (.3cm);
		\draw[fill=white] (0,0) circle (.29cm);
		\begin{scope}
		    	\clip (0,0) circle (.3cm);
		    	\foreach \x in {-.9,-.8,...,.3}
				\draw[line width=1 pt] (\x,-.3) -- (\x+.6,.3);
	  	\end{scope}
	  	
	  	% Neutrino loop
	  	\begin{scope}
	  		\clip (2,-2) rectangle (0,0);
		  	\draw[ postaction={decorate},
	decoration={markings,mark=at position .8 with {\arrow[draw=black]{>}}}] (0.9,0) ellipse (-0.6 and 0.3);
		  	\node at (0.8, -0.5) {$\nu$};
	  	\end{scope}
	  	\begin{scope}
	  		\clip (2, 2) rectangle (0,0);
		  	\draw[ postaction={decorate},
	decoration={markings,mark=at position .8 with {\arrow[draw=black]{>}}}] (0.9,0) ellipse (0.6 and -0.3);
		  	\node at (1, 0.5) {$\nu$};
	  	\end{scope}
	  	
	  	%Unitarity cut:
	  	\draw[dashed, thin] (0.9, -2) -- (0.9, 2);
	  	\node at (0.9, -2) [left] {s channel};
			
	  	\begin{scope}[shift={(1.8, 0)}]
		  	% B out:
			\draw[fermionbar](30:1.5) -- (30:0.3cm);
				\node at (30:1.7) {$B$};
			\begin{scope}[shift={(-35:-4pt)}]
				\draw[fermionbar](30:1.5) -- (30:0.3cm);
			\end{scope}
			\begin{scope}[shift={(-35:4pt)}]
				\draw[fermionbar](30:1.5) -- (30:0.3cm);
				\draw (30:0.3) -- (30:0);
			\end{scope}
			
			%\bar B out:
			\draw[fermion](-30:1.5) -- (-30:0.3cm);
				\node at (-30:1.7) {$ B$};
			\begin{scope}[shift={(35:-4pt)}]
				\draw[fermion](-30:1.5) -- (-30:0.3cm);
			\end{scope}
			\begin{scope}[shift={(35:4pt)}]
				\draw[fermion](-30:1.5) -- (-30:0.3cm);
				\draw (-30:0.3) -- (-30:0);
			\end{scope}
			
	  		% B blob:
			\draw[fill=black] (0,0) circle (.3cm);
			\draw[fill=white] (0,0) circle (.29cm);
			\begin{scope}
			    	\clip (0,0) circle (.3cm);
			    	\foreach \x in {-.9,-.8,...,.3}
					\draw[line width=1 pt] (\x,-.3) -- (\x+.6,.3);
		  	\end{scope}
	  	\end{scope}
	 \end{tikzpicture}
	 \caption{}
	 \label{fig:ABABs}
	 \end{subfigure}
	 \caption{Feynman diagrams for the neutrino-pair mediated (a)
           $AB\to AB$ scattering and (b) $A\bar A \to B\bar B$
           scattering. The labels in the figures denote the fields
           which describe the particles in the process.}
\end{figure}

Let's apply this relation to our process. We are interested in
calculating the absorptive part of the $AB \to AB$ amplitude mediated
by a neutrino-pair, so we need to do a $t$-channel unitarity cut of
the diagram in Fig.\ref{fig:ABABt}. Therefore, we should write
Eq.(\ref{eq:unit}) for the crossed process $A\bar A \to B\bar B$,
Fig.\ref{fig:ABABs}, with a $\nu \nu$ intermediate
state\footnote{Since the intermediate state is a fermionic one, there
  should be a spin sum. However, only left-handed neutrinos exist, so
  in this case it is not necessary.},
\begin{equation}
	\text{Im} \left\{  \bra{B\bar B} T \ket{A\bar A} \right\} = \frac{1}{2} \int \frac{\d^3 k_1}{(2\pi)^3 2E_{k_1}}\,\frac{\d^3 k_1}{(2\pi)^3 2E_{k_1}}\, \bra{\nu(k_1)\bar \nu(k_2)} T \ket{B\bar B}^*\bra{\nu(k_1) \bar \nu(k_2)}  T  \ket{A\bar A}\,.
\end{equation}

Dropping the $(2\pi)^4 \delta^{(4)}(p_f-p_i)$ global factor from both
sides, this equation becomes
\begin{equation} \begin{aligned}
	&\text{Im} \left\{  {\cal M}(A\bar A \to B\bar B) \right\} =\\
	&\hspace{1.25cm}=\frac{1}{2} \int \frac{\d^3 k_1}{(2\pi)^3 2E_{k_1}}\,\frac{\d^3 k_2}{(2\pi)^3 2E_{k_2}}\, (2\pi)^4 \delta^{(4)}(k_1+k_2-p_i) {\cal M}(B\bar B\to\nu \bar \nu)^*{\cal M}(A\bar A\to\nu \bar\nu)\,.
\end{aligned} \end{equation}

Finally, we can write this expression in an explicitly Lorentz
invariant manner,
\begin{gather}
	\boxed{ \begin{aligned} \text{Im} &\left\{  {\cal M}(A\bar A \to B\bar B) \right\} =\\
	&\hspace{0.75cm} = \frac{1}{2} \int \frac{\d^4 k_1}{(2\pi)^3}\, \delta(k_1^2)\,\frac{\d^4 k_2}{(2\pi)^3}\, \delta(k_2^2)\, (2\pi)^4 \delta^{(4)}(k_1+k_2-p_i) {\cal M}(B\bar B\to\nu \bar \nu)^*{\cal M}(A\bar A\to\nu \bar\nu) \end{aligned}}
	\raisetag{-0.125cm}
	\label{eq:ImMABAB}
\end{gather}

\newpage

\newpage\null\thispagestyle{empty}\newpage
\thispagestyle{newstyle}
\section{Low-Energy Contact Interaction}
\label{sec:Leff}

The weak interactions of fermions, charged and neutral currents, are
described by the Lagrangian densities \cite{pichEW, verde}
\begin{subequations} \begin{align}
	\Lagr_\text{CC} &=  -\frac{e}{2\sqrt{2} \sin\theta_W} \left\{ W^\dagger_\mu \left[ \bar u_i \gamma^\mu \left( 1-\gamma_5 \right) V_{ij}\, d_j + \bar \nu_i \gamma^\mu \left( 1-\gamma_5 \right) e_i\right]+ \text{h.c.} \begin{aligned} \mbox{}\\ \mbox{} \end{aligned} \right\}\,,\\
\nonumber  &\hspace{10cm} (i,j = 1^\text{st},\, 2^\text{nd},\, 3^\text{rd} \text{ gen.}) \\
	\Lagr_\text{NC} &= -e A_\mu \, Q_j \bar \psi_j \, \gamma^\mu \, \psi_j  - \frac{e}{4 \sin\theta_W \cos\theta_W}\, Z_ \mu \, \bar \psi_j \, \gamma^\mu \, \left( g_{V_j} - g_{A_j} \gamma_5 \right) \psi_j\\
\nonumber	&\equiv \Lagr_\text{QED}  + \Lagr_\text{Z} \,,  \hspace{7.5cm} (\psi_j= u, d, \nu_e, e...)
\end{align} \end{subequations}
where $\theta_W$ is the weak mixing angle.

For any elementary particle, the weak neutral couplings are given by
\begin{equation}
	g_V = 2T_3 - 4Q\sin^2\theta_W, \hspace{3cm} g_A = 2T_3\,,
	\label{eq:weakcharges}
\end{equation}
where $T_3$ is the third component of weak isospin and $Q$ is the
electric charge. The electroweak charges of the SM fermions are
written in Table \ref{tab:Q}.

\begin{table}[b]
	\centering
	\caption{Electroweak charges of the Standard Model fermions. The index $i=1,2,3$ labels the three generations, so that $u_1 = u,\, u_2 = c,\, u_3 = t...$}
	\label{tab:Q}
\begin{tabular}{ c | c c r }
	Particle	&$Q$			&$g_V$					&$g_A$\\ \hline \\
	$u_i$ 		&\nicefrac{2}{3}&$\hspace{10pt}1-\frac{8}{3}\sin^2\theta_W$	&1\\
	$d_i$ 		&\nicefrac{-1}{3}		&$-1+\frac{4}{3}\sin^2\theta_W$	&-1\\
	$\nu_i$	&0			&$1$						&1 \\
	$e_i$		&-1			&$-1+4\sin^2\theta_W$			&-1\\
%	p [uud]	&1			&$1-4\sin^2\theta_W$	&1\\
%	n [udd]	&0			&$-1$				&-1\\
\end{tabular}
\end{table}

We are interested in calculating the potential associated to a process
at low energy, where the limit $|q^2|\ll M_W^2, M_Z^2$ is valid, so
now we'll focus in obtaining the low-energy effective interactions
from the above Lagrangians.

\subsection{Effective charged 
current couplings}

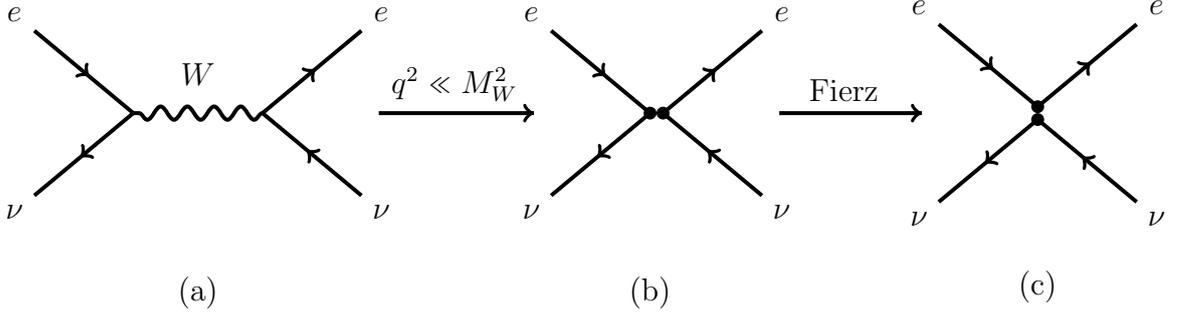
\begin{figure}[t]
	\centering
	\begin{tikzpicture}[line width=1.5 pt, scale=1.7]
		%Particles in:
		\draw[fermionbar] (-140:1)--(0,0);
			\node at (-140:1.2) {$\nu$};
		\draw[fermion] (140:1)--(0,0);
			\node at (140:1.2) {$e$};

		%Propagator
		\draw[vector] (0:1)--(0,0);
			\node at (.5,.3) {$W$};
		\node at (0.5, -1.4) {(a)};

		%Particles out:
		\begin{scope}[shift={(1,0)}]
			\draw[fermion] (-40:1)--(0,0);
				\node at (-40:1.2) {$\nu$};
			\draw[fermionbar] (40:1)--(0,0);
				\node at (40:1.2) {$e$};	
		\end{scope}
		
		%Change of diagram
		\begin{scope}[shift={(1.9,0)}]
			\clip (0, -1) rectangle (1.2, 1);
			\draw[fermion] (0,0) -- (2.18, 0);
				\node at (0,0)[above right] {$q^2 \ll M_W^2$};
		\end{scope}
		
		%Particles in:
		\begin{scope}[shift={(4,0)}]
			\draw[fermionbar] (-140:1)--(0,0);
				\node at (-140:1.2) {$\nu$};
			\draw[fermion] (140:1)--(0,0);
				\node at (140:1.2) {$e$};
			\fill[black] (0,0) circle (0.05);
			
			\node at (0, -1.4) {(b)};
		\end{scope}
		
		%Particles out:
		\begin{scope}[shift={(4.1,0)}]
			\draw[fermion] (-40:1)--(0,0);
				\node at (-40:1.2) {$\nu$};
			\draw[fermionbar] (40:1)--(0,0);
				\node at (40:1.2) {$e$};	
			\fill[black] (0,0) circle (0.05);
		\end{scope}
		
		%Change of diagram
		\begin{scope}[shift={(5,0)}]
			\clip (0, -1) rectangle (1.1, 1);
			\draw[fermion] (0,0) -- (2, 0);
				\node at (0.5,0)[above] {Fierz};
		\end{scope}
		
		%Particles up:
		\begin{scope}[shift={(7,0.05)}]
			\draw[fermionbar] (40:1)--(0,0);
				\node at (40:1.2) {$e$};	
			\draw[fermion] (140:1)--(0,0);
				\node at (140:1.2) {$e$};
			\fill[black] (0,0) circle (0.05);
			
			\node at (0, -1.4) {(c)};
		\end{scope}
		
		%Particles down:
		\begin{scope}[shift={(7,-0.05)}]
			\draw[fermionbar] (-140:1)--(0,0);
				\node at (-140:1.2) {$\nu$};
			\draw[fermion] (-40:1)--(0,0);
				\node at (-40:1.2) {$\nu$};
			\fill[black] (0,0) circle (0.05);
		\end{scope}
	\end{tikzpicture}
	\caption{Tree-level Feynman diagrams for the
          $\bar \nu e \to \bar \nu e$ scattering corresponding to (a)
          the Standard Model charged current Lagrangian, (b) the
          effective low-energy Lagrangian obtained integrating out the
          $W$ degrees of freedom and (c) this last Lagrangian after
          Fierz reordering the fields and writing the interaction
          currents as flavor-diagonal.}
	\label{fig:CCeff}
\end{figure}

We are describing neutrino scattering against an aggregate of matter,
so only the $\nu_e$-$e$ charged current contributes to the
scattering. Therefore, the only two terms of the interaction
Lagrangian which are interesting to our process are
\begin{equation}
	\Lagr_\text{CC} = M_W^2 W_\mu^\dagger W^\mu +  W_\mu^\dagger \, \bar \nu_e\,  \Gamma^\mu\, e + W_\mu \, \bar e \, \Gamma^\mu \, \nu_e\,,
	\label{eq:Lcc}
\end{equation}
where
\begin{equation*}
	\Gamma^\mu \equiv -\frac{e}{2\sqrt{2}\sin\theta_W}\gamma^\mu (1-\gamma_5) \vspace{0.4cm}
\end{equation*}
and we also wrote the kinetic term of the $W_\mu$ field.

In order to calculate the effective Lagrangian, we integrate the
$W_\mu$ degrees of freedom out of the Lagrangian using its equations
of motion,
\begin{equation}
	0 = \frac{\partial \Lagr_\text{CC}}{\partial W_\mu^\dagger} = M_W^2 W^\mu + \bar \nu_e\,  \Gamma^\mu\, e\,,
\end{equation}
so
\begin{equation}
	W_\mu = -\frac{1}{M_W^2}\,\bar \nu_e\,  \Gamma_\mu\, e =
        \frac{e}{2\sqrt{2} M_W^2 \sin\theta_W}\,\bar \nu_e\,  \gamma_\mu\,(1-\gamma_5)\, e \,.
\end{equation}

Putting this relation into Eq.(\ref{eq:Lcc}) one easily gets
\begin{equation}
	\Lagr^\text{eff}_\text{CC} = - \frac{G_F}{\sqrt{2}} \left[ \bar \nu_e\,  \gamma^\mu\,(1-\gamma_5)\, e \right] \left[ \bar e \,  \gamma_\mu\,(1-\gamma_5)\, \nu_e \right]\,,
\end{equation}
where the Fermi constant is given by
$\frac{G_F}{\sqrt{2}} = \frac{e^2}{8M_W^2 \sin^2\theta_W}$.

It is convenient to write this Lagrangian as flavour diagonal---as
shown in Fig.\ref{fig:CCeff}---, so that we can add both CC and NC
Lagrangians. In order to do so, we use the Fierz identity
(\ref{eq:Fierz_fields}) and the relation
$\gamma_\mu \gamma_\nu \gamma^\mu = -2 \gamma_\nu$ to write
\begin{equation}
	\Lagr^\text{eff}_\text{CC} = -\frac{G_F}{\sqrt{2}} \left[ \bar \nu_e\,  \gamma^\mu\,(1-\gamma_5)\, \nu_e \right] \left[ \bar e \,  \gamma_\mu\,(1-\gamma_5)\, e \right]\,.
	\label{eq:CCeff}
\end{equation}
%This whole process is represented in Fig.\ref{fig:CCeff}.

\subsection{Effective neutral current couplings}

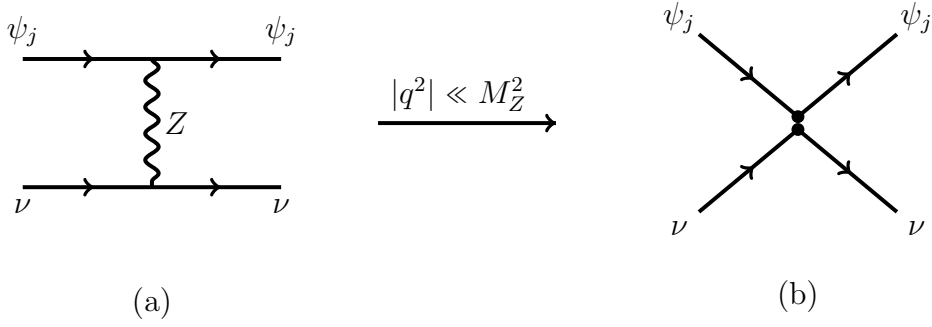
\begin{figure}[t]
	\centering
	\begin{tikzpicture}[line width=1.5 pt, scale=1.7]
		%Particles up:
		\begin{scope}[shift={(0,0.5)}]
			\draw[fermion] (180:1)--(0,0);
				\node at (180:1) [above] {$\psi_j$};	
			\draw[fermionbar] (0:1)--(0,0);
				\node at (0:1)[above] {$\psi_j$};
		\end{scope}

		%Propagator
		\draw[vector] (0, 0.5)--(0,-0.5);
			\node at (0,0) [right] {$Z$};
		\node at (0, -1.4) {(a)};

		%Particles down:
		\begin{scope}[shift={(0,-0.5)}]
			\draw[fermion] (180:1)--(0,0);
				\node at (180:1)[below] {$\nu$};
			\draw[fermionbar] (0:1)--(0,0);
				\node at (0:1)[below] {$\nu$};
		\end{scope}
		
		%Change of diagram
		\begin{scope}[shift={(1.75,0)}]
			\clip (0, -1) rectangle (1.375, 1);
			\draw[fermion] (0,0) -- (2.5, 0);
				\node at (0,0)[above right] {$\left| q^2 \right| \ll M_Z^2$};
		\end{scope}
				
		%Particles up:
		\begin{scope}[shift={(5,0.05)}]
			\draw[fermionbar] (40:1)--(0,0);
				\node at (40:1.2) {$\psi_j$};	
			\draw[fermion] (140:1)--(0,0);
				\node at (140:1.2) {$\psi_j$};
			\fill[black] (0,0) circle (0.05);
			
			\node at (0, -1.4) {(b)};
		\end{scope}
		
		%Particles down:
		\begin{scope}[shift={(5,-0.05)}]
			\draw[fermion] (-140:1)--(0,0);
				\node at (-140:1.2) {$\nu$};
			\draw[fermionbar] (-40:1)--(0,0);
				\node at (-40:1.2) {$\nu$};
			\fill[black] (0,0) circle (0.05);
		\end{scope}
	\end{tikzpicture}
	\caption{Tree-level Feynman diagrams for the $\psi \nu \to
          \psi \nu $
          scattering corresponding to (a) the Standard Model neutral
          current Lagrangian and (b) the effective low-energy
          Lagrangian obtained integrating out the $Z$ degrees of
          freedom.}
	\label{fig:NCeff}
\end{figure}

In this case, the interesting Lagrangian to our process is
\begin{equation}
	\Lagr_\text{NC} = \frac{1}{2} M_Z^2 Z_\mu Z^\mu - \frac{e}{4\sin\theta_W \cos\theta_W} \,Z_ \mu \, \bar \psi_j \, \gamma^\mu \, \left( g_{V_j} - g_{A_j} \gamma_5 \right) \psi_j\,,
	\label{eq:Lnc}
\end{equation}
where $j = u, d, e, \nu_e, \nu_\mu, \nu_\tau$.

As in the previous section, we integrate out the $Z$ degrees of
freedom using its equations of motion,
\begin{equation}
	0 = \frac{\partial \Lagr_\text{NC}}{\partial Z_\mu} = M_Z^2 Z^\mu - \frac{e}{4\sin\theta_W \cos\theta_W} \,\bar \psi_j \, \gamma^\mu \, \left( g_{V_j} - g_{A_j} \gamma_5 \right) \psi_j\,,
\end{equation}
so
\begin{equation}
	Z_\mu =  \frac{e}{4\sin\theta_W \cos\theta_W M_Z^2} \,\bar \psi_j \, \gamma_\mu \, \left( g_{V_j} - g_{A_j} \gamma_5 \right) \psi_j\,.
\end{equation}

Putting this relation in Eq.(\ref{eq:Lnc}) we get
\begin{equation}
	\Lagr_\text{NC}^\text{eff} = -\frac{G_F}{2\sqrt{2}}\left[ \bar \nu\,  \gamma^\mu\,(1-\gamma_5)\, \nu \right] \left[ \bar \psi_j \,  \gamma_\mu\,\left( g_{V_j} - g_{A_j} \gamma_5 \right) \psi_j \right]\,,
	\label{eq:NCeff}
\end{equation}
as is represented in Fig.\ref{fig:NCeff}.

%%%%%%%%%		 WEAK FLAVOUR CHARGES		 %%%%%%%%%%%%%

\subsection{Low-energy effective Lagrangian for matter particles}
Let us now consider some aggregate of matter $A$. In the scattering
process $A \nu \to A\nu$ at low energy, the neutrino can interact with
the three ``elementary'' particles which matter is formed
with---electrons, protons and neutrons.

We can consider that nucleons are point-like Dirac particles because
the scattering happens at low energy---i.e. the neutrino is like a
large scale probe, so it cannot resolve the structure of nucleons. The
vector current is conserved, so both the electric charge $Q$ and the
weak vector charge $g_V$ of the nucleon are the sum of its valence
quarks' charges,
\begin{minipage}{0.45\textwidth}
	\begin{align}
	\nonumber	\hspace{2cm}  Q_p &=1\,,\\
	\nonumber	Q_n &=0\,,
	\end{align}
\end{minipage}
\hfill
\begin{minipage}{0.5\textwidth}
	\begin{align}
	\nonumber g_V^p &= 1-4\sin^2\theta_W\,, \hspace{1cm} \\
		g_V^n &= -1\,.	\label{eq:QWN}
	\end{align}
\end{minipage}\vspace{0.3cm}

On the other hand, the axial current is not conserved, so this
argument does not apply to the weak axial charge of the nucleon. In
fact, Eq.(\ref{eq:weakcharges}) shows that the axial coupling is
independent of the electric charge---it only depends on the weak
isospin coupled to the $W_\mu^3$ boson. Therefore, it can be expected
due to weak isospin\footnote{In fact, for the first generation of
  quarks, weak and strong isospin coincide.} symmetry that the weak
neutral axial coupling at low momentum transfer, $q^2\to 0$, is the
same as the coupling to the $W_\mu^\pm$ mediated charge current
responsible of the $n\to p$ process, $g_A = 1.2723\pm0.0023$
\cite{PDG}.

Taking all of this into account, the Lagrangian describing the
$A \nu\to A \nu$ interaction has three terms, related to the processes
\[ \begin{array}{ c l}
	\hspace{4cm} \nu_e + e \longrightarrow \nu_e + e\,, &\\
	\hspace{4cm}\nu_i + e \longrightarrow \nu_i + e\,, &\hspace{3cm} (i = \mu, \tau)\\
	\hspace{4cm}\nu_j + N \longrightarrow \nu_j + N\,. &\hspace{3cm} (j = e, \mu, \tau)
\end{array} \]

The first one is mediated by both charged and neutral
currents. Therefore, we have to add the Lagrangians (\ref{eq:CCeff})
and (\ref{eq:NCeff}), so we get
\begin{align}
\nonumber	\Lagr_1 &= -\frac{G_F}{\sqrt{2}} \left[ \bar \nu_e\,
                          \gamma^\mu\,(1-\gamma_5)\, \nu_e \right]
                          \left[ \bar e \,  \gamma_\mu\,(1-\gamma_5)\,
                          e \right] - \\
\nonumber	&\hspace{2cm}- \frac{G_F}{2\sqrt{2}}\left[ \bar \nu_e\,  \gamma^\mu\,(1-\gamma_5)\, \nu_e \right] \left[ \bar e \,  \gamma_\mu\,\left( g_V^e - g_A^e \gamma_5 \right) e \right] = \\
	&= -\frac{G_F}{2\sqrt{2}}\left[ \bar \nu_e \,  \gamma^\mu\,(1-\gamma_5)\, \nu_e \right] \left[ \bar e \,  \gamma_\mu\,\left( \tilde g_V^e - \tilde g_A^e \gamma_5 \right) e \right]\,,  \label{eq:L1}
\end{align}
where we have defined
\begin{align}
	\nonumber \tilde g_V^e &= 2 + g_V^e = 1 + 4 \sin^2\theta_W\,, \\
	\tilde g_A^e &= 2 + g_A^e = 1\,.	\label{eq:QWe}
\end{align}

The second and third ones are only mediated by neutral currents, so
they are described by the Lagrangian (\ref{eq:NCeff}),
\begin{align}
	\Lagr_2 &= -\frac{G_F}{2\sqrt{2}}\left[ \bar \nu_i \,  \gamma^\mu\,(1-\gamma_5)\, \nu_i \right] \left[ \bar e \,  \gamma_\mu\,\left( g_V^e - g_A^e \gamma_5 \right) e \right]\,, \hspace{3cm} (i = \mu, \tau)
\label{eq:L2}\\
	\Lagr_3 &= -\frac{G_F}{2\sqrt{2}}\left[ \bar \nu_j \,  \gamma^\mu\,(1-\gamma_5)\, \nu_j \right] \left[ \bar N \,  \gamma_\mu\,\left( g_V^N - g_A^N \gamma_5 \right) N \right]\,. \hspace{2.35cm} \begin{aligned} &(j = e, \mu, \tau)\\ &(N = p, n) \end{aligned}
	\label{eq:L3}
\end{align}

\begin{figure}[t]
	\centering
	\begin{tikzpicture}[line width=1.5 pt, scale=1.7]
		%Particles in:
		\begin{scope}[shift={(-0.05,0)}]
			\draw[fermionbar] (135:1)--(0,0);
				\node at (135:1.1)[left] {$\psi_j$};	
			\draw[fermion] (-135:1)--(0,0);
				\node at (-135:1.1)[left] {$\psi_j$};
			\fill[black] (0,0) circle (0.05);
			
			\node at (-0.25,0)[left] {$\frac{G_F}{2\sqrt{2}}\, \gamma_\mu \left(g_V^j - g_A^j \gamma_5 \right)$};
		\end{scope}
		
		%Particles out:
		\begin{scope}[shift={(0.05,0)}]
			\draw[fermion] (45:1)--(0,0);
				\node at (45:1.1)[right] {$\nu_i$};
			\draw[fermionbar] (-45:1)--(0,0);
				\node at (-45:1.1)[right] {$\nu_i$};
			\fill[black] (0,0) circle (0.05);
			
			\node at (0.25,0)[right] {$ \gamma^\mu \left( 1 -  \gamma_5 \right)$};
		\end{scope}
		
		\node at (3, 0.3)[right]{$\psi_j = e, p, n$};
		\node at (3, -0.3)[right]{$\nu_i = \nu_e, \nu_\mu, \nu_\tau$};
	\end{tikzpicture}
	\caption{Fundamental vertex of the effective low-energy
          Lagrangian (\ref{eq:L123}). The couplings $g_V,\, g_A$
          depend on both the neutrino flavor and which is the charged
          fermion, as discussed in the text.}
	\label{fig:Leff}
\end{figure}
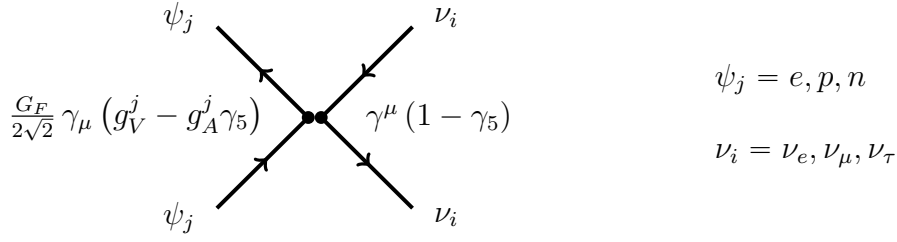

With all this information, we finally have our whole interaction
Lagrangian $\Lagr = \Lagr_1 + \Lagr_2 + \Lagr_3$, which is
\begin{equation} \boxed{\begin{aligned}
	\Lagr = -\frac{G_F}{2\sqrt{2}} &\left\{ \left[ \bar \nu_e \,  \gamma^\mu\,(1-\gamma_5)\, \nu_e \right] \left[ \bar e \,  \gamma_\mu\,\left( \tilde g_V^e - \tilde g_A^e \gamma_5 \right) e \right] + \begin{aligned} \mbox{}\\ \mbox{} \end{aligned} \right. \\
	&+ \left[ \bar \nu_i \,  \gamma^\mu\,(1-\gamma_5)\, \nu_i \right] \left[ \bar e \,  \gamma_\mu\,\left( g_V^e - g_A^e \gamma_5 \right) e \right] + \hspace{3.5cm} (i = \mu, \tau) \\
	&\left.+ \left[ \bar \nu_j \,  \gamma^\mu\,(1-\gamma_5)\, \nu_j \right] \left[ \bar N \,  \gamma_\mu\,\left( g_V^N - g_A^N \gamma_5 \right) N \right] \begin{aligned} \mbox{}\\ \mbox{} \end{aligned} \right\} \hspace{2.75cm} \begin{aligned} &(j = e, \mu, \tau)\\ &(N = p, n) \end{aligned}
\end{aligned}} 
\label{eq:L123}
\end{equation}

All fundamental vertices of this Lagrangian have the same structure,
as is represented in Fig.\ref{fig:Leff}. As a check, the couplings we
obtained here are (indeed) the same ones stated in \cite{verde}.

%%%%%%%%%		 AMPLITUDES  			%%%%%%%%%%%%%

\newpage
\newpage\null\thispagestyle{empty}\newpage
\thispagestyle{newstyle}
\section{Neutrino-Matter Scattering}
\label{sec:amplitude}

% Once the interaction Lagrangian is written, we can focus on
% calculating ${\cal M}(B\nu \to B\nu )^*{\cal M}(A\nu\to A\nu)$, a
% quantity needed to get the absorptive part of the $AB \to AB$
% amplitude, as Eq.(\ref{eq:ImM}) states \textcolor{red}{After
% crossing! Can I cross so easily?}. In fact, it will prove useful to
% work in the Breit reference frame, defined by $q^0 = 0$, so that
%\begin{equation}
%	\text{Im} \left\{  {\cal M}( AB \to AB) \right\} = \frac{1}{32 \pi^2} \int \,\d\Omega \,{\cal M}(B\nu \to B\nu )^*{\cal M}(A\nu\to A\nu )\,.
%\end{equation}
Once the interaction Lagrangian is written, we can focus on
calculating the scattering amplitude between an aggregate of matter
and a neutrino, ${\cal M}(A\nu \to A\nu)$. For simplicity, $A$ can be
understood as a molecule, composed of $Z_A$ protons and electrons and
$N_A$ neutrons---we'd better study electrically neutral systems, with
the same number of protons and electrons, because any net-charge
electric interaction would be much stronger than the weak interaction
we're looking for.

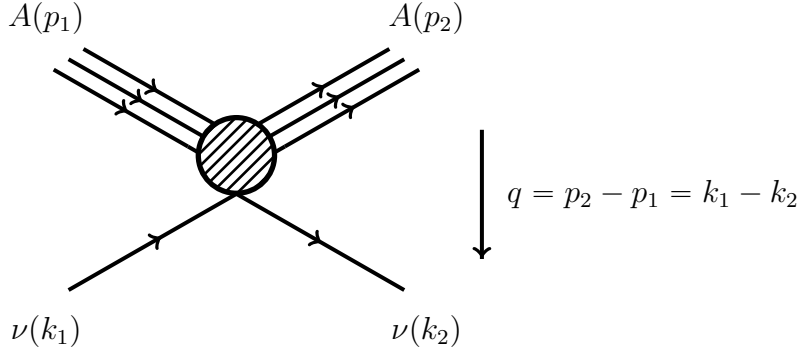
\begin{figure}[t]
	\centering
	\begin{tikzpicture}[line width = 1.5pt, scale = 1.7]
		% A in:
		\draw[fermion](150:1.5) -- (150:0.3cm);
			\node at (150:1.7)[above] {$A(p_1)$};
		\begin{scope}[shift={(35:4pt)}]
			\draw[fermion](150:1.5) -- (150:0.3cm);
		\end{scope}
		\begin{scope}[shift={(35:-4pt)}]
			\draw[fermion](150:1.5) -- (150:0.3cm);
			\draw (150:0.3) -- (150:0);
		\end{scope}
		% A out:
		\draw[fermionbar](30:1.5) -- (30:0.3cm);
			\node at (30:1.7)[above] {$A(p_2)$};
		\begin{scope}[shift={(-35:-4pt)}]
			\draw[fermionbar](30:1.5) -- (30:0.3cm);
		\end{scope}
		\begin{scope}[shift={(-35:4pt)}]
			\draw[fermionbar](30:1.5) -- (30:0.3cm);
			\draw (30:0.3) -- (30:0);
		\end{scope}
		% A blob:
		\draw[fill=black] (0,0) circle (.3cm);
		\draw[fill=white] (0,0) circle (.29cm);
		\begin{scope}
		    	\clip (0,0) circle (.3cm);
		    	\foreach \x in {-.9,-.8,...,.3}
				\draw[line width=1 pt] (\x,-.3) -- (\x+.6,.3);
	  	\end{scope}
	  	%Neutrinos
	  	\begin{scope}[shift={(0, -0.3)}]
			\draw[fermion] (-150:1.5) -- (0,0);
				\node at (-150:1.7)[below]{$\nu (k_1)$};
			\draw[fermionbar](-30:1.5) -- (0,0);
				\node at (-30:1.7)[below]{$\nu (k_2)$};
		\end{scope}
		%Momentum transfer
	  	\begin{scope}[shift={(1.9,-0.3)}]
	  		\begin{scope}
	  			\clip (-1,1) rectangle (0.5,-0.5);
	  			\draw[fermion] (0,0.5) -- (0, -1.31);
	  		\end{scope}
	  		\node at (0.1,0)[right]{$q = p_2 - p_1 = k_1 - k_2$};
		\end{scope}
	  \end{tikzpicture}
	  \caption{Lowest order Feynman diagram for the $A \nu \to
            A\nu$ scattering in the low-energy effective weak theory.}
	  \label{fig:AnuAnu}
\end{figure}

As shown in Fig.\ref{fig:AnuAnu}, the process $A \nu \to A\nu$ is
described by an elementary vertex of the interaction Lagrangian
(\ref{eq:L123}). The different terms of this Lagrangian show
explicitly that the coupling neutrino-matter must depend on the flavor
of the neutrino, so we will consider them separately.

\subsection{Electron neutrino}
In the $A (p_1) \, \nu_e (k_1) \to A(p_2) \, \nu_e(k_2) $ case, the amplitude is determined by
\begin{equation} 
	i\,T_{\nu_e} = \bra{A \nu_e} i \int \d^4 x \left[ \Lagr_1(x) +\Lagr_2(x) +\Lagr_3(x) \right]  \ket{A \nu_e} \equiv i\,T_{\nu_e}^{(1)} + i\,T_{\nu_e}^{(2)} +i\, T_{\nu_e}^{(3)}\,,
\end{equation}
which we can calculate separately. The contribution of the first term is
\begin{align}
	\nonumber T_{\nu_e}^{(1)} &= \bra{A \nu_e} \int \d^4 x  \, \frac{-G_F}{2\sqrt{2}}\left[ \bar \nu_e \,  \gamma^\mu\,(1-\gamma_5)\, \nu_e \right] \left[ \bar e \,  \gamma_\mu\,\left( \tilde g_V^e - \tilde g_A^e \gamma_5 \right) e \right] \ket{A \nu_e} = \\
	\nonumber &= -\frac{G_F}{2\sqrt{2}}\int \d^4 x  \, \bra{\nu_e}  \bar \nu_e \,  \gamma^\mu\,(1-\gamma_5)\, \nu_e  \ket{\nu_e} \, \bra{A} \bar e \,  \gamma_\mu\,\left( \tilde g_V^e - \tilde g_A^e \gamma_5 \right) e  \ket{A} \equiv \\
	&\equiv -\frac{G_F}{2\sqrt{2}}\int \d^4 x  \,  j^\mu_{\nu_e}(x) J_\mu^{(1)}(x) \,.	\label{eq:Tnue}
\end{align}

Since neutrinos are elementary particles, the leptonic current is the
usual
\begin{align}
	\nonumber j^\mu (x) &= \bra{\nu_e (k_2) }  \bar \nu_e (x) \,  \gamma^\mu\,(1-\gamma_5)\, \nu_e (x)  \ket{\nu_e (k_1)} = \\
	&=\left[ \bar u (k_2)\, \gamma^\mu \,(1-\gamma_5) \, u(k_1) \right]\, e^{-i(k_1-k_2)x}\,. 
\end{align} 

On the other hand, the molecular current matrix element will be
\begin{equation}
	J_\mu^{(1)} (x) = e^{-i(p_1-p_2)x}\, J_\mu^{(1)} = e^{-i(p_1-p_2)x}\, \bra{A (p_2)} \bar e (0) \gamma_\mu \left(\tilde g_V^e - \tilde g_A^e \gamma_5 \right) e (0) \ket{ A (p_1)} \,.
\end{equation}

Since we're looking for a low-energy coherent interaction, it's
interesting to analyze separately the different terms in $J_\mu$:
\begin{itemize}
  \item $\gamma^0$ is a scalar quantity, related to the matrix element
    of $e^\dagger e$, which is the number operator, so its
    contribution is coherent.
  \item $\gamma^0 \gamma_5$ is a pseudo-scalar quantity, so its matrix
    element is related to $\vec \sigma \vec q / M$, where
    $\vec \sigma$ is the spin of $A$, $M$ its mass and
    $\vec q = \vec p_1 - \vec p_2$. Since this contribution depends on
    $\vec \sigma$, it's not coherent. Also, any contribution of the
    form $\vec q / M$ gives a relativistic correction to the
    potential, so this is another reason why we can ignore this term.
  \item $\vec \gamma$ is a polar vector, so its matrix element must be
    proportional to $\vec q/M$. Again, this is a relativistic
    correction we won't consider.
  \item $\vec \gamma \gamma_5$ is an axial vector, directly related to
    the spin of the particle, so this contribution is not coherent.
\end{itemize}

Therefore, the coherent contribution to the molecular current is given
by
%\footnote {Aside for a normalization factor which we must omit in calculating the potential, as discussed in section \ref{sec:Potentials}.}
\begin{align}
	\nonumber J_0^{(1)} (x) &= \bra{A (p_2)} \bar e (x)\, \tilde g_V^e \gamma^0 e (x) \ket{ A (p_1)} = \tilde g_V^e \bra{A (p_2)} e^\dagger (x) \, e (x) \ket{ A (p_1)} =\\
	\nonumber &= \tilde g_V^e \bra{A (p_2)} \left[ \int \d^4y \ket{y}\bra{y} \right] e^\dagger (x) \, e (x) \left[ \int \d^4z \ket{z}\bra{z} \right] \ket{ A (p_1)} =\\
	\nonumber &=\tilde g_V^e \int \d^4y\, \d^4 z\, e^{ip_2y}\, e^{-ip_1z}\, \bra{A (y)} e^\dagger (x) \, e (x) \ket{ A (z)} = \\
	\nonumber &=\tilde g_V^e \int \d^4y\, \d^4 z\, e^{ip_2y}\, e^{-ip_1z}\, \delta^{(4)}(x-y) \, \delta^{(4)}(x-z) Z_A = \\
	&= Z_A\, \tilde g_V^e\, e^{-i(p_1-p_2)x}\,,
\end{align}
where we have inserted two Closure Relations,
\begin{equation}
	I = \int \d^4x \ket{x}\bra{x}\,,
\end{equation}
and we have taken into account the fact that $e^\dagger (x) e(x) $ is
the electron number operator at $x$.

Even though $J_0$ is the only relevant component of $J_\mu$, it is
convenient to keep a relativistic framework---later we'll consider the
non-relativistic limit. Therefore, the $T$ matrix element
(\ref{eq:Tnue}) is
\begin{equation}
	T_{\nu_e}^{(1)} = (2\pi)^4 \delta^{(4)}(q + k_1-k_2) \times \frac{-G_F}{2\sqrt{2}} \, J^{(1)}_\mu \left[ \bar u (k_2)\, \gamma^\mu \,(1-\gamma_5) \, u(k_1) \right] \,,
\end{equation}
where $q \equiv p_1-p_2$ and $J_0^{(1)} = Z_A \tilde g_V^e$---this
last equality, and the following giving $J_0$ values, must be
understood as the coherent contribution to $J_0$ given by the number
operator of the particle constituents.

Analogously,
\begin{align}
	T_{\nu_e}^{(2)} &=0\,,\\
	T_{\nu_e}^{(3)} &= (2\pi)^4 \delta^{(4)}(q + k_1-k_2) \times \frac{-G_F}{2\sqrt{2}} \, J^{(3)}_\mu \left[ \bar u (k_2)\, \gamma^\mu \,(1-\gamma_5) \, u(k_1) \right] \,.
\end{align}
where $J_0^{(3)} = Z_A  g_V^p + N_A g_V^n$.

Adding all contributions and dropping the $(2\pi)^4 \delta(p_i-p_f)$
factor, we get
\begin{equation}
	{\cal M}(A\nu_e \to A\nu_e) = -\frac{G_F}{2\sqrt{2}} \, J_{A, \mu}^e \left[ \bar u (k_2)\, \gamma^\mu \,(1-\gamma_5) \, u(k_1) \right]\,,
	\label{eq:Me}
\end{equation}
where $J_{A, \mu}^e$ is the molecular current in the scattering with
an electron neutrino. Using the weak charges from (\ref{eq:QWN}) and
(\ref{eq:QWe}),

\noindent\begin{minipage}{0.45\textwidth}
	\begin{align}
	\nonumber	\hspace{2cm}  g_V^p &=1 - 4 \sin^2\theta_W\,,\\
	\nonumber	g_V^n &= -1\,,\\
	\nonumber
	\end{align}
\end{minipage}
\hfill
\begin{minipage}{0.5\textwidth}
	\begin{align}
	\nonumber	g_V^e &= -1+4\sin^2\theta_W = - g_V^p\,, \hspace{1cm} \\
	\nonumber	\tilde g_V^e &= 2 + g_V^e = 2- g_V^p\,,\\
	\nonumber
	\end{align}
\end{minipage}\vspace{0.25cm}\\
we find $J_{A, 0}^e = 2Z_A-N_A$.  At this level, we remind the reader
that the first term comes from charged current interaction with
electrons, while the second one comes from neutral currents with
neutrons---neutral currents with protons and electrons cancel out.

\subsection{Muon and tau neutrino}
The muon and tau flavors have the same contribution to the Effective
Lagrangian, so the scattering amplitudes for the processes
$A \nu_\mu \to A \nu_\mu$ and $A \nu_\tau \to A \nu_\tau$ must be the
same. Therefore, we can consider both of them simultaneously and
calculate
\begin{equation}
	i\,T_{\nu_j} = \bra{A \nu_j} i \int \d^4 x \left[ \Lagr_1(x) +\Lagr_2(x) +\Lagr_3(x) \right]  \ket{A \nu_j} \equiv  i\,T_{\nu_j}^{(2)} +i\, T_{\nu_j}^{(3)}\,,
\end{equation}
where $j=\mu, \tau$ and $\Lagr_1$ does not contribute because it only
has electron neutrinos. Following the same steps than in the previous
section, we get
\begin{align}
	T_{\nu_j}^{(2)} &= (2\pi)^4 \delta^{(4)}(q + k_1-k_2) \times \frac{-G_F}{2\sqrt{2}} \, J_{A, \mu}^{(2)} \left[ \bar u (k_2)\, \gamma^\mu \,(1-\gamma_5) \, u(k_1) \right] \,,\\
	T_{\nu_j}^{(3)} &= (2\pi)^4 \delta^{(4)}(q + k_1-k_2) \times \frac{-G_F}{2\sqrt{2}} \,  J_{A, \mu}^{(3)} \left[ \bar u (k_2)\, \gamma^\mu \,(1-\gamma_5) \, u(k_1) \right] \,,
\end{align}
where $J_{A, 0}^{(2)} = Z_A g_V^e$ and
$J_{A, 0}^{(3)} = Z_A g_V^p + N_A g_V^n $.

Finally, dropping the $(2\pi)^4 \delta(p_i-p_f)$ factor, we can write
\begin{equation}
	{\cal M}(A\nu_j \to A\nu_j) = -\frac{G_F}{2\sqrt{2}} \,J_{A, \mu}^j \left[ \bar u (k_2)\, \gamma^\mu \,(1-\gamma_5) \, u(k_1) \right]\,,  \hspace{2cm} \left( j=\mu, \tau \right)\,,\\
	\label{eq:Mmutau}
\end{equation}
where $J_{A, 0}^j = -N_A$. As before, the neutral current interactions
for electrons and protons cancel each other in neutral (of electric
charge) matter.

\subsection{The weak flavor charge of aggregate matter}
Up to this point, we have calculated the amplitudes of the processes
described by all fundamental vertices of our Lagrangian, so it's
convenient to sum up our results and analyze them. In order to do
that, it's useful to compare with well-known theories.

Let's consider QED. In this theory, a process described by the
fundamental vertex would involve two fermions and a photon. If we take
the photon on-shell and drop external-leg fermion factors, the $iM$
would be

\noindent \begin{minipage}{0.9\textwidth}
\centering
	\hspace{0.1\textwidth}
	\begin{tikzpicture}[line width = 1.5pt, scale = 1.7]
		\draw[fermion] (-0.87, 0.5) -- (0,0);
			\node at (-0.87, 0.5)[above left]{$\psi$};
		\draw[fermionbar] (0.87, 0.5) -- (0,0);
			\node at (0.87, 0.5)[above right]{$\psi$};
		\draw[vector] (0,0) -- (0, -1);
			\node at (0, -1)[above right]{$\gamma (k)$};
		\node at (2, -0.25){$ = -i e Q \gamma^\mu \epsilon_\mu (k)\,,$};
	\end{tikzpicture}
\end{minipage}
\begin{minipage}{0.1\textwidth}
	\begin{equation}
		\mbox{}
	\end{equation}
\end{minipage} \vspace{0.5cm} \\
where $Q$ is the electric charge of the fermion field. Due to vector
current conservation, this vertex does also apply to non-fundamental
particles, which have an electric charge equal to the sum of its
constituents' charges---and the amplitude would be this charge times
the coupling $e$.

This same behavior appears in weak interactions. The only
flavor-diagonal interaction is the one mediated by neutral currents,
with the fundamental vertex

\noindent\begin{minipage}{0.9\textwidth}
\centering
	\hspace{0.1\textwidth}
	\begin{tikzpicture}[line width = 1.5pt, scale = 1.7]
		\draw[fermion] (-0.87, 0.5) -- (0,0);
			\node at (-0.87, 0.5)[above left]{$\psi$};
		\draw[fermionbar] (0.87, 0.5) -- (0,0);
			\node at (0.87, 0.5)[above right]{$\psi$};
		\draw[vector] (0,0) -- (0, -1);
			\node at (0, -1)[right]{$Z (k)$};
		\node at (1, -0.425)[right]{$ = -i \frac{e}{4\sin\theta_W \cos\theta_W} \left\{  \left( 2T_3-4Q\sin^2\theta_W \right) \gamma^\mu - 2T_3 \gamma^\mu \gamma_5 \right\} \epsilon_\mu (k)\,.$};
	\end{tikzpicture}
\end{minipage}
\begin{minipage}{0.1\textwidth}
	\begin{equation}
		\mbox{}
	\end{equation}
\end{minipage}\vspace{0.5cm}

As is thoroughly discussed in \cite{TFG}, the vector current
conservation allows us to talk about a weak charge of the fermion
field $\psi$, which is $Q_W = 2T_3-4Q\sin^2\theta_W$, such that the
weak charge of a composed particle is the sum of its constituents', as
happens with electric charge. However, we can't talk about the axial
coupling as a charge, since the axial current is not conserved.

Once the concept of a weak charge has been introduced, we see that the
vector part of this amplitude has the same structure as the QED one, a
coupling times a charge times a mediating-particle external-leg
factor. According to this idea, we can expect our amplitudes to have
this structure too.

Indeed, both Eqs.(\ref{eq:Me}) and (\ref{eq:Mmutau}) can be written
(in the non-relativistic limit) as

\noindent \begin{minipage}{0.9\textwidth}
\centering
	\hspace{0.1\textwidth}
	\begin{tikzpicture}[line width = 1.5pt, scale = 1.7]
		%Move the whole picture
		\begin{scope}
			\clip  (-1,-1) rectangle (1.2,1.2);
		\end{scope}
		% A in:
		\draw[fermion](150:1.5) -- (150:0.3cm);
			\node at (150:1.7)[left] {$A$};
		\begin{scope}[shift={(35:4pt)}]
			\draw[fermion](150:1.5) -- (150:0.3cm);
		\end{scope}
		\begin{scope}[shift={(35:-4pt)}]
			\draw[fermion](150:1.5) -- (150:0.3cm);
			\draw (150:0.3) -- (150:0);
		\end{scope}
		% A out:
		\draw[fermionbar](30:1.5) -- (30:0.3cm);
			\node at (30:1.7)[right] {$A$};
		\begin{scope}[shift={(-35:-4pt)}]
			\draw[fermionbar](30:1.5) -- (30:0.3cm);
		\end{scope}
		\begin{scope}[shift={(-35:4pt)}]
			\draw[fermionbar](30:1.5) -- (30:0.3cm);
			\draw (30:0.3) -- (30:0);
		\end{scope}
		% A blob:
		\draw[fill=black] (0,0) circle (.3cm);
		\draw[fill=white] (0,0) circle (.29cm);
		\begin{scope}
		    	\clip (0,0) circle (.3cm);
		    	\foreach \x in {-.9,-.8,...,.3}
				\draw[line width=1 pt] (\x,-.3) -- (\x+.6,.3);
	  	\end{scope}
	  	%Neutrinos
	  	\begin{scope}[shift={(0, -0.3)}]
			\draw[fermion] (-150:1.5) -- (0,0);
				\node at (-150:1.7)[left]{$\nu_i (k_1)$};
			\draw[fermionbar](-30:1.5) -- (0,0);
				\node at (-30:1.7)[right]{$\nu_i (k_2)$};
		\end{scope}
		%Feynman rule:
		\node at (3, -0.3){$ = i \frac{G_F}{2\sqrt{2}}\, J_{A, 0}^i\, \left[ \bar u (k_2) \gamma^0 (1-\gamma_5)  u(k_1) \right]\,,$};
	  \end{tikzpicture}\end{minipage}
\begin{minipage}{0.1\textwidth}
	\begin{equation}
		\mbox{}
	\end{equation}
\end{minipage}\vspace{0.5cm}\\
where $Q_{W,A}^i \equiv J_{A, 0}^i$ is the weak charge of the
aggregate of matter $A$. It depends on the flavor of the neutrino, so
we can speak of three weak flavor charges of aggregate matter, which
are given by
\begin{align}
	 \nonumber Q_{W,A}^e &= 2Z_A-N_A\,,\\
	 Q_{W,A}^\mu = Q_{W,A}^\tau &= - N_A\,.
	 \label{eq:weakflavorcharges}
\end{align}
Eqs.(\ref{eq:weakflavorcharges}) state the fact that, whereas
aggregate matter is neutral of electric charge, it is not neutral of
weak charges!

\begin{figure}[t]
	\centering
	\includegraphics[width=0.65\textwidth]{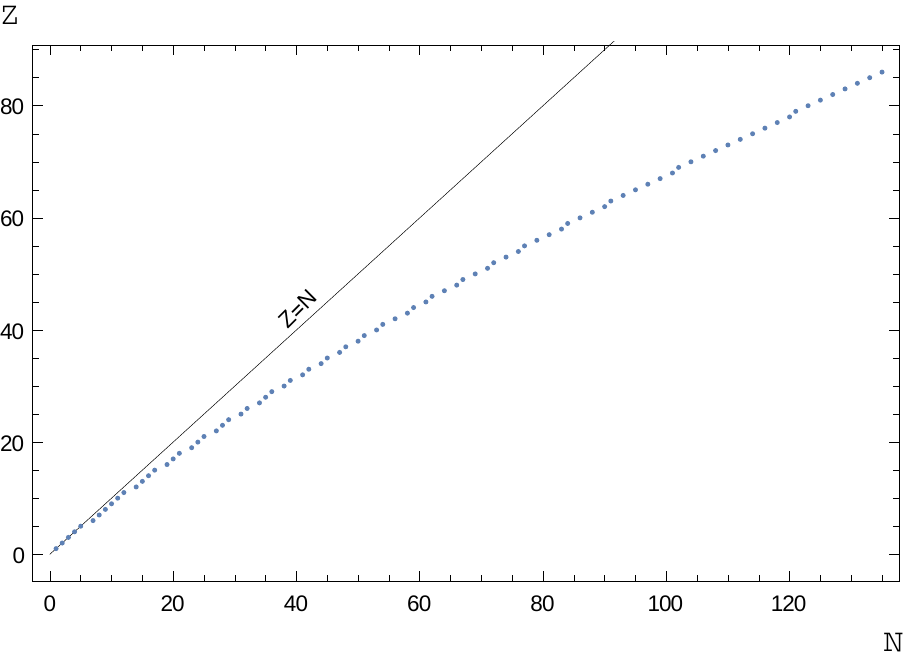}
	\caption{The valley of stability, composed of the pairs of
          $(Z,N)$ for all elements, according to the semi-empirical
          mass formula, Eq.(\ref{eq:semimass}).}
	\label{fig:valle}
\end{figure}

\begin{figure}[h!]
	\centering
	\includegraphics[width=0.65\textwidth]{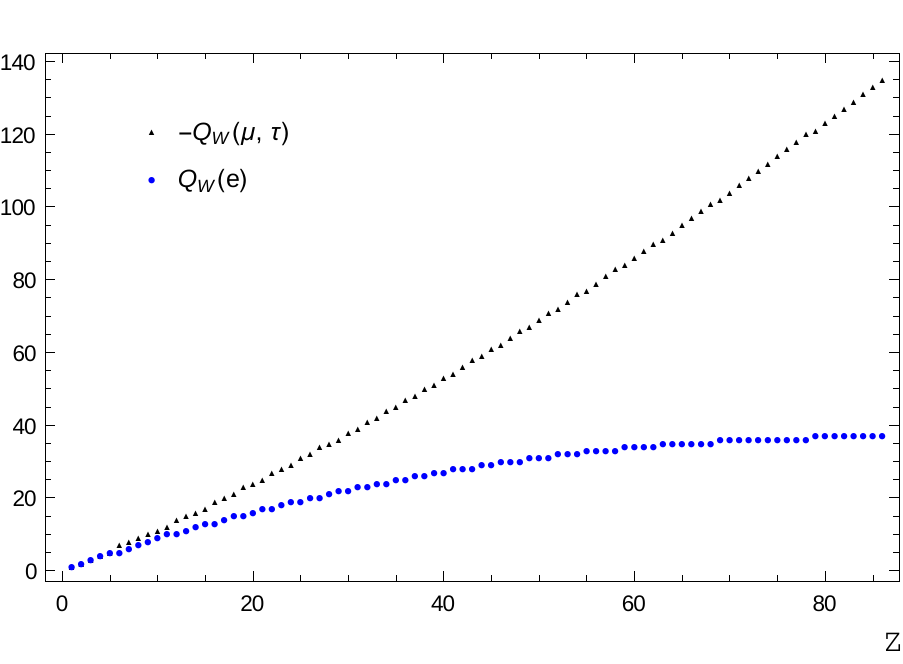}
	\caption{Weak flavor charges of the elements with $(Z,N)$ in
          the valley of stability, Fig.\ref{fig:valle}. Beware a minus
          sign in the $\mu, \tau$ flavor charges.}
	\label{fig:qweak}
\end{figure}

It's interesting to analyze the value of those charges for ``normal''
matter. In order to do that, we'll look at stable nuclei. According to
the semi-empirical mass formula \cite{krane}, the $(Z,N)$ values of
stable nuclei are related by
\begin{equation}
	Z \approx \frac{A}{2+0.0157 A^{2/3}}\,,
	\label{eq:semimass}
\end{equation}
where $A \equiv Z+N$, as is represented in Fig.\ref{fig:valle}. Using
those pairs of values, the weak charges of each element (neutral atom)
are represented in Fig.\ref{fig:qweak}, where we see that the electron
neutrino weak charge is always positive, while the muon and tau
neutrino charges are always negative. The weak charge of aggregate
matter is obtained from Fig.\ref{fig:qweak} by multiplying by the
number of the constituent atoms.
\newpage

\newpage
\newpage\null\thispagestyle{empty}\newpage
\thispagestyle{newstyle}
\section{Long-Range Weak Interaction Potential}
\label{sec:weakpot}
After analyzing the $A\nu \to A\nu$ scattering amplitude, we can focus
on calculating the interaction potential. We'll begin using
Eq.(\ref{eq:ImMABAB}) to determine the absorptive part of the
$AB\to AB$ amplitude. After that, we'll use Eq.(\ref{eq:VIm}) to
obtain the potential.

\subsection{Absorptive part of $AB \to AB$ at low $t$}

In order to get the absorptive part of the scattering amplitude, we
need to compute the crossed quantity
Im$\{ {\cal M}(A \bar A \to B \bar B) \}$, as written in
Eq.(\ref{eq:ImMABAB}). Therefore, we need to cross the amplitude we
calculated in the previous section, from
\begin{equation*}
	{\cal M}(A\nu_i \to A\nu_i) = -\frac{G_F}{2\sqrt{2}} \,J_{A, \mu}^i \left[ \bar u (k_2)\, \gamma^\mu \,(1-\gamma_5) \, u(k_1) \right]
\end{equation*}
to
\begin{equation}
	{\cal M}(A \bar A\to\nu_i \bar \nu_i) = -\frac{G_F}{2\sqrt{2}} \, \tilde J_{A, \mu}^i \left[ \bar u (k_2)\, \gamma^\mu \,(1-\gamma_5) \, v(k_1) \right]\,,
\end{equation}
where $\tilde J_\mu$ is the crossed molecular current, which still
satisfies $\tilde J^i_0 = Q^i_W$ in the non-relativistic limit.

According to Eq.(\ref{eq:ImMABAB}), the absorptive part of
${\cal M}(A\bar A\to B \bar B)$ is determined by the quantity
${\cal M}(A \bar A \to \nu_i \bar \nu_i) {\cal M}^*(B \bar B \to \nu_i
\bar \nu_i)$,
which we can now evaluate. From now on we'll work in a simplified
case, assuming that neutrinos are massless\footnote{The non-vanishing
  mass of neutrinos will affect the behavior of the potential at the
  longest range---its implications will be announced in Section
  \ref{sec:prospects}. Prospects.}, so
\begin{align}
	\nonumber {\cal M}(A \bar A \to \nu_i \bar \nu_i)&{\cal M}^*(B \bar B \to \nu_i \bar \nu_i)  =\\
	\nonumber &= \frac{G_F^2}{8}\, \tilde J_{A, \mu}^i \, \tilde J_{B, \nu}^i \left[ \bar u (k_2)\, \gamma^\mu \,(1-\gamma_5) \, v(k_1) \right] \left[ \bar v(k_1)\, \gamma^\nu \,(1-\gamma_5) \, u(k_2) \right] = \\
	\nonumber &= \frac{G_F^2}{8}\, \tilde Z^i_{\mu\nu} \text{ Tr}\left[ \slashed{k_1} \gamma^\nu (1-\gamma_5) \slashed{k_2} \gamma^\mu (1-\gamma_5) \right] = \\
	\nonumber &= \frac{G_F^2}{4}\, \tilde Z^i_{\mu\nu} \text{ Tr}\left[ \slashed{k_1} \gamma^\nu  \slashed{k_2} \gamma^\mu (1-\gamma_5)\right] = \\
	&=  G_F^2\, \tilde Z^i_{\mu\nu} \, \left[ k_1^\mu k_2^\nu + k_1^\nu k_2^\mu - g^{\mu\nu}(k_1 k_2) + a^{\mu\nu} \right]\,,
\end{align}
where we defined
$\tilde Z^i_{\mu\nu} \equiv \tilde J_{A, \mu}^i \, \tilde J_{B,
  \nu}^i$
and $a^{\mu\nu}$ is some antisymmetric tensor which we will no longer
consider because it vanishes in the non-relativistic limit, where the
only relevant component is $\mu = \nu = 0$.

Considering the contributions of the three neutrino flavors, the
absorptive part is
\begin{align}
	\nonumber\text{Im} \left\{  {\cal M}\right.&\left.\hspace{-3pt}(A\bar A \to B \bar B) \right\} = \\
	\nonumber &= \frac{G_F^2}{8\pi^2}\,\left( \sum_f \tilde Z^f_{\mu\nu} \right)\, \int \d^4 k_1\, \delta(k_1^2)\, \delta(k_2^2)\,\left[ k_1^\mu k_2^\nu + k_1^\nu k_2^\mu - \frac{1}{2} sg^{\mu\nu} \right] = \\
	\nonumber &= \frac{G_F^2}{8\pi^2}\,\left( \sum_f \tilde Z^f_{\mu\nu} \right)\, \int \d^4 k_1\, \delta(k_1^2)\, \delta(k_2^2)\,\left[ -2k_1^\mu k_1^\nu + \left( k_1^\mu q^\nu + k_1^\nu q^\mu \right) - \frac{1}{2} s g^{\mu\nu} \right] = \\
	\nonumber &= \frac{G_F^2}{8\pi^2}\,\left( \sum_f \tilde Z^f_{\mu\nu} \right)\, \frac{\pi}{2}\,\left[ -\frac{2}{3}\left( q^\mu q^\nu - \frac{1}{4} t g^{\mu\nu} \right) + \frac{1}{2} \left( q^\mu q^\nu + q^\nu q^\mu \right) - \frac{1}{2} s g^{\mu\nu}  \right] = \\
	& = \frac{G_F^2}{24\pi}\,\left( \sum_f \tilde Z^f_{\mu\nu} \right)\, \left[\,q^\mu q^\nu - s g^{\mu\nu} \,\right]\,, \label{eq:Imtensor}
\end{align}
where we used $k_2 = q - k_1$ in the second line and all integrals
needed in the third line are stated in \cite{sucher67}---we
demonstrate them in Appendix \ref{sec:intsuch}.

As seen, the tensor structure of Eq.(\ref{eq:Imtensor}) is transverse, a requirement which any quantity built from conserved currents must satisfy. We can cross this result back to the $t-$channel applying $s \to t$, so that
\begin{equation}
	\text{Im} \left\{ {\cal M}(AB \to AB) \right\} = \frac{G_F^2}{24\pi}\,\left( \sum_f Z^i_{\mu\nu} \right)\, \left[ q^\mu q^\nu - t g^{\mu \nu} \right]\,.
\end{equation}

Now it's easy to evaluate the non-relativistic limit. As discussed
before, the only relevant component of the molecular current for
coherent interactions is the scalar contribution to $J_0$, so we can
take
\begin{equation}
	\text{Im} \left\{ M(AB \to AB) \right\} = \frac{G_F^2}{24\pi}\,\left( \sum_f Q_{W,A}^f \, Q_{W,B}^f \right)\, \left[ \left(q^0\right)^2 - t \right]\,.
\end{equation}

Besides, we are looking for a long-range interaction, so
$q^0 \approx 0$ and
\begin{equation}
	\text{Im} \left\{ M(AB \to AB) \right\} = - \frac{G_F^2}{24\pi}\,\left( \sum_f Q_{W,A}^f \, Q_{W,B}^f \right)\,  t \,,
\label{eq:ImMAB}
\end{equation}
where
\begin{equation}
	\sum_f Q_{W,A}^f \, Q_{W,B}^f = (2Z_A-N_A)(2Z_B-N_B) +2N_AN_B\,.
	%&= 4Z_AZ_B -2(Z_AN_B+Z_BN_A)+N_AN_B\,.
	\label{eq:weakcoupling}
\end{equation}

\begin{figure}[t]
	\centering
	\includegraphics[width=0.65\textwidth]{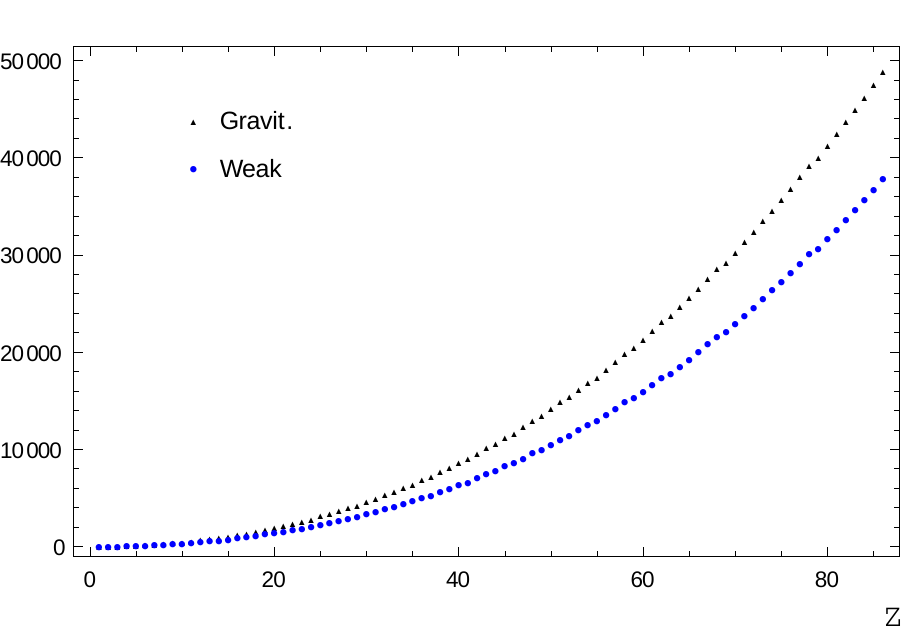}
	\caption{Weak coupling $\sum Q_W^f Q_W^f$, which is written in
          Eq.(\ref{eq:weakcoupling}), for the elements of the valley
          of stability (Fig.\ref{fig:valle}), each one interacting
          with itself. The gravitational coupling
          \nicefrac{$M^2$}{$m_p^2$} $\approx (Z+N)^2$, neglecting
          binding energies, is also represented.}
	\label{fig:qqweak}
\end{figure}

As Eq.(\ref{eq:ImMAB}) shows, all the flavor dependence of the
absorptive part---and, therefore, of the potential---in the limit of
massless neutrinos is factorized in the weak charges.

As we saw in Fig.\ref{fig:qweak}, all the stable elements have the
same sign for the weak charges, $Q_W^e > 0$ and $Q_W^{\mu,
  \tau}<0$.
This implies that, for any pair of elements---and therefore for any
pair of molecules---this coupling has a positive sign, so the
resulting force will have the same character---whether repulsive or
attractive---for any pair of aggregates of matter. We'll find out
which of those two cases is the right one in the next section.

Let's see the behavior of this quantity for some cases. As we did in
the previous section, we'll consider only stable nuclei. Since this is
an interaction, we have to choose sets of two elements---we'll
consider the interaction of each element with itself. In
Fig.\ref{fig:qqweak} we show the quantity $\sum Q^f_{W,A}Q^f_{W,A}$
for the element $A$ as a function of the atomic number $Z_A$ for the
stable nuclei.

In the same Figure we compare the weak coupling with the gravitational
one. It's seen that both of them increase with the number of particles
of the systems interacting, but they scale differently, even when the
binding energy is neglected. That means that our coherent weak
interaction could introduce a deviation from the Equivalence
Principle, as was announced in \cite{taup}.

\subsection{Neutrino-pair exchange potential}
After obtaining this result, the only remaining step in the
calculation of the interaction potential is computing the integral
(\ref{eq:VIm}), with the branching point at $t_0 = 0$ (for massless
neutrinos). Using the Im$\{M\}$ obtained in Eq.(\ref{eq:ImMAB}),
\begin{equation}
	V(r) =  \frac{G_F^2}{24 \pi}\,\left( \sum_f Q_{W,A}^f \, Q_{W,B}^f \right) \frac{1}{4\pi^2 r}\, \int_0^\infty \, \d t\, t \,  e^{-\sqrt{t} \,r}\,,
	\label{eq:intV}
\end{equation}
so we have to evaluate this integral. A primitive $P(t)$ is calculated
in Appendix \ref{sec:AppI1}, so we obtain the integral (say $I$) using Barrow's Rule,
$I = P(t\to \infty) - P(t\to 0)$. If we assume $r \neq 0$ when
calculating the limits---which is valid, since we are looking for a
long-range interaction---this contribution to the potential is
\begin{equation}
	I = \frac{12}{r^4} \,.
\end{equation}

With this result, we can finally write
\begin{equation}
	\boxed{  \hspace{0.25cm}\begin{aligned} \mbox{}\\ \mbox{} \end{aligned}  V(r) = \frac{G_F^2}{8\pi^3}\,\left( \sum_f Q_{W,A}^f \, Q_{W,B}^f \right)  \frac{1}{r^5} \hspace{0.25cm}}\,,
	\label{eq:V(r)}
\end{equation}
which has an associated force given by
\begin{equation}
	\vec{F}(r) = - \grad V(r) = \frac{5 G_F^2}{8 \pi^3}\,\left( \sum_f Q_{W,A}^f \, Q_{W,B}^f \right)  \frac{\vec{\hat r}}{r^6}\,,
\end{equation}
where $\hat r$ is the radial unit vector.

We have obtained a long-range interaction which is repulsive for
ordinary matter, since the weak coupling is always positive---as we
showed in Fig.\ref{fig:qqweak}. This is a difference with the
gravitational force that we can use to distinguish them---and we can
also look for the deviations from the Equivalence Principle we
discussed in the previous section.

The other interactions that appear in electrically neutral systems are
the residual electromagnetic Van der Waals ($\sim r^{-7}$) or
Casimir-Polder ($\sim r^{-8}$) forces, which have a lower range
(larger inverse power law) than our weak force, even though they're
stronger. It would be interesting to find systems with low
electromagnetic momenta, so that this interactions became weaker, as
the ones described in \cite{isotech}.

\newpage
\thispagestyle{newstyle}
\section{Conclusions}
\label{sec:conclusions}

We began this work reviewing the relation between the description of
an interaction process in the framework of a Quantum Field Theory and
in terms of an interaction potential. As shown in Section
\ref{sec:Potentials}, the Feynman amplitude of an elastic scattering
process and the effective potential describing this interaction are
related, in Born approximation, by a Fourier Transform. This is the
result one would expect, taking into account that $M(q^2)$ describes
the interaction process in momentum space, while $V(r)$ describes the
interaction in position space.

Another interesting detail that stems from the fact that we are
calculating a long-range interaction is that we needn't calculate the
whole amplitude of the process in order to determine the
potential. This is due to the fact that $M(q^2)$ at $|q^2|\to \infty$
gives short-range contributions, so the potential is determined by
Im$\{ M(q^2) \}$ through an unsubtracted dispersion
relation. Therefore, we could use the unitarity relation from Section
\ref{sec:unitarity} to avoid the calculation of $M(q^2)$ (a $1-$loop
quantity) and compute a tree-level process instead. Besides, the fact
that we were interested in the low-energy limit allowed us to work in
the framework of an effective theory where Charged Currents and
Neutral Currents could be written in the form of a contact
interaction, so we worked with only one interaction vertex.

The results from Sections \ref{sec:Potentials}-\ref{sec:Leff} made it
clear that the effective potential was determined by the amplitude of
the process $A \nu \to A\nu$, which we calculated in Section
\ref{sec:amplitude}. Although this was a quite straightforward
calculation, it gave rise to a very interesting concept---the weak
flavor charge of aggregate matter. Indeed, the coupling of bulk matter
to a neutrino is proportional to $G_F$ with a charge that depends on
the flavor of the neutrino,
\[
	Q_W^e = 2Z-N\,, \hspace{3cm} Q_W^\mu = Q_W^\tau = -N\,.
\]
These are the weak flavor charges for electrically neutral
matter---the case we are interested in, since any non-zero electric
charge would produce an electromagnetic interaction much stronger than
our weak interaction.

This amplitude was the last ingredient needed to compute the effective
potential, which gives raise to the repulsive force
\[
	\vec F = \frac{5 G_F^2}{8 \pi^3} \left( \sum_f Q_{W,A}^f Q_{W,B}^f \right) \frac{\vec{\hat r}}{r^6}\,,
\]
where all flavor dependence is in the weak charges. This is the
coherent contribution to the force, which we obtained from the vector
charge $J^0$, proportional to $\gamma^0$. The first correction to this
result would come from the spin dependent contribution to $\vec J$,
that comes from $\vec \gamma \gamma_5$. The other two contributions to
the current, proportional to $\gamma^0\gamma_5$ and $\vec \gamma$,
give relativistic corrections $\sim \frac{1}{M}$.

In the long-range regime we are looking at, there are two other
important interactions: residual electromagnetic interactions and
gravitation. For ordinary molecules, Van der Waals forces are much
stronger than our weak interaction at short distances, so it would be
interesting to look for a system where the first electromagnetic
moments are zero. In the case of gravitation, there are two traits of
this weak interaction that can help to distinguish between them in an
experiment: this force is repulsive---while gravitation is
attractive---and its charge is proportional to the number of particles
but not to their mass, so it would produce a signal that deviates from
the Equivalence Principle.

In any case, joining the previous ideas with the recent development of
atomic traps \cite{trap}---not ionic traps---can be the key to observe
this interaction in an experiment.

\newpage
\thispagestyle{newstyle}
\section{Prospects}
\label{sec:prospects}
The long-range potential obtained in this work, Eq.(\ref{eq:V(r)}), is
valid and of interest for distances between nanometers and
microns. The short-distance limit comes from the requirement of having
neutral (of electric charge) systems of aggregate matter, while the
long-distance limit is imposed by a non-vanishing value of the
absolute mass of the neutrino---indeed, the range of this interaction
for neutrinos of $m\sim 0.1$ eV is of the order of
\[
	R \sim \frac{\hbar c}{mc^2} = \frac{197 \text{ MeV fm}}{0.1 \text{ eV}} \sim 10^9 \text{ fm} = 1\, \mu\text{m}\,.
\]
In this region, the effective potential will become of Yukawa type
instead of the inverse power law. We can get a first idea on the
dependence of the potential with $m$ changing slightly this work's
result. If we had integrated Eq.(\ref{eq:intV}) from a branching point
at $t_0 =~4m^2$, the potential would have been
\[
	V(r) = \frac{G_F^2}{8 \pi^3}\,\left( \sum_f Q_{W,A}^f \, Q_{W,B}^f \right) \left( \frac{1}{r^5} + \frac{2m}{r^4} + \frac{2m^2}{r^3} + \frac{4m^3}{3r^2}\right) e^{-2mr}\,,
\]
which depends on $m$ not only in the Yukawa exponential, but also in
the preceding inverse power terms\footnote{Of course, the computation
  of the potential at finite $m$ is not so trivial---the mass has to
  be included in the absorptive part of the amplitude---, but it
  serves for illustrating the kind of changes that will occur.}.
% Another interesting change would come from neutrino oscillation, so
% that we don't expect the flavor dependence of the potential to be
% factorized with a non-zero neutrino mass.

The neutrino mass dependence of the effective potential in the
long-range behavior opens novel directions in the study of the most
interesting pending questions on neutrino properties: absolute
neutrino mass (from the range), flavor dependence and mixing (from the
weak charges in the interaction) and, hopefully, with two neutrino
exchange, the exploration of the most crucial open problem in neutrino
physics: whether neutrinos are Dirac or Majorana particles.

The study of these problems will be the subject of my immediate future
research work. On the one hand, it's necessary to calculate the form
of this interaction with a finite mass for the neutrino. In fact, two
calculations are needed: for Dirac neutrinos and for Majorana
neutrinos. On the other hand, the collaboration with the experimental
groups involved in neutral traps will be initiated this summer during
my stay at CERN, in order to find out whether the low electromagnetic
interacting systems mentioned in our Conclusions could be implemented.

%%%%%%%%%%%%%%%%%%%%%%%%%%%%%%%%%%%%%%%%%%%%%%
%								APÉNDICES
%%%%%%%%%%%%%%%%%%%%%%%%%%%%%%%%%%%%%%%%%%%%%%
\newpage
\newpage\null\thispagestyle{empty}\newpage
\addtocontents{toc}{\protect\newpage} % Adds \newpage in "\tableofcontents"
\addcontentsline{toc}{part}{Appendices}
\appendix
\thispagestyle{newstyle}

\noindent \textbf{\huge Appendices}

\section{Notations and Conventions}
\label{sec:appA}
\subsection{Units}
This work is written using the Natural System of Units, where
$\hbar = c = k_B = 1$. We describe electromagnetic quantities with the
Heavyside system, $\epsilon_0 = \mu_0 = 1$, so that the fine-structure
constant is given by $\alpha = e^2/4\pi \approx 1/137$.
 
\subsection{Relativity end Tensors}
We define the Minkowsi metric tensor with signature $(+, -, -, -)$, as
\begin{equation}
	g_{\mu \nu} = g^{\mu \nu} = \left( \begin{array}{c c c c}
	1&0&0&0\\
	0&-1&0&0\\
	0&0&-1&0\\
	0&0&0&-1\\
	\end{array} \right)\,,
\end{equation}
so that any 4-vector can be written as
$x^\mu = \left( x^0, \vec x \right)$. We also use the Einstein
Summation Convention, so scalar products can be written as
$x\cdot p = x^\mu p_\mu = g_{\mu \nu} x^\mu p^\nu = x^0 p^0 - \vec x
\cdot \vec p$.
We can also write
$x_\mu = g_{\mu \nu} x^\nu = \left( x^0, -\vec x \right)$, and the
derivative operator is
$\partial_\mu \equiv \frac{\partial}{\partial x^\mu} =
\left( \partial_t, \grad \right)$.

% We define de D'Alembertian as
% $\Box \equiv \partial^2 = \partial_\mu \partial^\mu = \partial_t^2 -
% \grad^2 $.

We use the totally antisymmetric
tensor %$\epsilon^{\alpha \beta \gamma \delta}$
with the convention $\epsilon^{0 1 2 3} = +1 = - \epsilon_{0123}$.

\subsection{Fourier Transforms}
We define Fourier Transforms so that all $2\pi$ factors are included
in the momentum integration,
%\begin{subequations} \begin{align}
%	f (x) &= \int \frac{\d^4 k}{(2\pi)^4}\, e^{-ikx}\, \tilde f(k)\,,\\
%	\tilde f(k) & = \int \d^4 x \, e^{ikx}\, f(x)\,.
%\end{align}\end{subequations}
\[\]
\noindent\begin{minipage}{0.5\textwidth}
%	\begin{center} \underline{4-dimensional FT:} \end{center}
	\underline{$4-$dimensional FT:}
	\begin{align}
	\nonumber f (x) &= \int \frac{\d^4 k}{(2\pi)^4}\, e^{-ikx}\, \tilde f(k)\,,\\
	\nonumber \tilde f(k) & = \int \d^4 x \, e^{ikx}\, f(x)\,,
	\end{align}
\end{minipage}
\noindent\begin{minipage}{0.5\textwidth}
%	\begin{center} \underline{3-dimensional FT:} \end{center}
	\underline{$3-$dimensional FT:}
	\begin{align}
	\nonumber f (\vec x) &= \int \frac{\d^3 k}{(2\pi)^3}\, e^{i\vec k \vec x}\, \tilde f(\vec k)\,,\\
	\tilde f(\vec k) & = \int \d^3 x \, e^{-i\vec k \vec x}\, f(\vec x)\,,
	\end{align}
\end{minipage}

Other $2\pi$ factors come from the following expression for the Dirac
delta,
\begin{equation}
	\int \d^4 x \,  e^{ikx} = (2\pi)^4 \delta^{(4)}(k) \,.
\end{equation}

\subsection{Diracology}
As is well known, Dirac gamma matrices are required to satisfy the
relations
\begin{equation}
	\{ \gamma^\mu, \gamma^\nu \} = 2g^{\mu \nu}\,, \hspace{3cm}
	[\gamma^\mu, \gamma^\nu] = -2i\sigma^{\mu \nu}.
\end{equation}
Also,
\begin{equation}
	(\gamma^0)^2 = - (\gamma^i)^2 = 1, \hspace{3cm} \gamma_\mu^\dagger = \gamma^0 \gamma_\mu \gamma^0.
\end{equation}

We define the fifth gamma matrix as
$\gamma_5 \equiv i \gamma^0 \gamma^1 \gamma^2 \gamma^3 = -
\frac{i}{4!} \epsilon_{\alpha \beta \gamma \delta}\, \gamma^\alpha
\gamma^\beta \gamma^\gamma \gamma^\delta $,
which satisfies $(\gamma_5)^2 = -1, \; \gamma_5^\dagger =
\gamma_5$. Therefore, the quirality projectors can be written as
\begin{equation}
	P_R = \frac{1+\gamma_5}{2}, \hspace{3cm} P_L = \frac{1-\gamma_5}{2}.
	\label{eq:PLR}
\end{equation}

Some useful contractions are
\begin{subequations} \begin{align}
	&\gamma^\mu \gamma_\mu = 4\,,\hspace{6.8cm}\\
	&\gamma^\mu \gamma^\nu \gamma_\mu = -2 \gamma^\nu,\\
	&\gamma^\mu \gamma^\alpha \gamma^\beta \gamma_\mu = 4 g^{\alpha \beta},\\
	&\gamma^\mu \gamma^\nu \gamma^\alpha \gamma^\beta \gamma_\mu = -2\gamma^\beta \gamma^\alpha \gamma^\nu.
\end{align} \end{subequations}

Some useful trace identities are
\begin{subequations} \begin{align}
	&\text{Tr}\left[ \gamma^\mu \gamma^\nu \right] = 4 g^{\mu \nu},\\
	&\text{Tr}\left[ \gamma^\mu \gamma^\nu \gamma_5 \right] = 0\,,\\
	&\text{Tr}\left[ \gamma^\mu \gamma^\nu  \gamma^\alpha \gamma^\beta  \right] = 4 \left( g^{\mu\nu} g^{\alpha\beta} - g^{\mu\alpha}g^{\nu\beta}+g^{\mu\beta}g^{\nu\alpha} \right), \\
	&\text{Tr}\left[ \gamma^\mu \gamma^\nu  \gamma^\alpha \gamma^\beta  \gamma_5 \right] = -4i \, \epsilon^{\mu\nu\alpha\beta},\\
	&\text{Tr}\left[ \gamma^{\mu_1}\gamma^{\mu_2}...\gamma^{\mu_{2k+1}} \right] = 0\,.
\end{align} \end{subequations}

For any $4-$vector $a^\mu$, we define $\slashed{a} \equiv \gamma_\mu a^\mu$.

\newpage
\thispagestyle{newstyle}
\section{Useful Relations}

\subsection{Fierz Identity}
Let us consider the Dirac-scalar quantity
\begin{equation}
	\left[ \bar u_1 A P_L u_2 \right] \left[ \bar u_3 P_R B u_4 \right]\,,
	\label{eq:buuuuu}
\end{equation}
where $A$ and $B$ are arbitrary matrices in Dirac space, $P_{L,R}$ are
the quirality projectors from Eq.(\ref{eq:PLR}) and the four $u_i$ are
Dirac spinors\footnote{If any of these spinors were a $v$ spinor,
  nothing in this section would change---the Identity would still hold.}.

The set of matrices
$\Gamma_i = \{ 1, \gamma_5, \gamma^\mu P_L, \gamma^\mu P_R,
\sigma^{\mu \nu} \}$ are a basis of Dirac space, so we can expand
\begin{equation}
	u_2 \bar u_3 = \sum_i \alpha_i \Gamma_i = \alpha_1 1 + \alpha_5 \gamma_5 + \alpha_L^\mu \gamma_\mu P_L + \alpha_R^\mu \gamma_\mu P_R + \alpha_S^{\mu \nu} \sigma_{\mu \nu}\,.
\end{equation}

Since we have this expansion between quiarilty projectors, we can
simplify
\begin{equation}
	P_L u_2 \bar u_3 P_R = P_L \left(  \sum_i \alpha_i \Gamma_i \right) P_R = \alpha_R^\mu \gamma_\mu P_R\,,
	\label{eq:exp}
\end{equation}
where we have used the relations
\[ 
	P_{L,R}^2 = P_{L,R}\,,\hspace{1cm}	P_{L,R} P_{R,L} = 0\,,\hspace{1cm}	P_{L,R}\gamma_\mu = \gamma_\mu P_{R,L}\,,\hspace{1cm}	 P_{L,R}\gamma_5 = \gamma_5 P_{L,R}
\]
to show that all other terms are zero.

We need to calculate the $\alpha_R^\mu$ coefficient, so we evaluate
the quantity
\begin{equation} 
	\text{Tr}[\gamma^\nu P_L u_2 \bar u_3] = \text{Tr}\left[\gamma^\nu P_L \left(  \sum_i \alpha_i \Gamma_i \right) \right] = \alpha_R^\mu \text{Tr}[\gamma^\nu P_L \gamma_\mu P_R] = 2 \alpha_R^\mu\,,
\end{equation}
which means that we can get the coefficient by computing
\begin{equation}
	\alpha_R^\mu = \frac{1}{2}\text{Tr}[\gamma^\mu P_L u_2 \bar u_3] = \frac{1}{4}\bar u_3 \gamma^\mu (1-\gamma_5) u_2 = \frac{1}{2} \bar u_3 \gamma^\mu P_L u_2\,.
\end{equation}

Putting this relation into Eq.(\ref{eq:exp}) one gets
\begin{equation}
	P_L u_2 \bar u_3 P_R =  \frac{1}{2} [\bar u_3 \gamma^\mu P_L u_2] \gamma_\mu P_R\,,
\end{equation}
so that Eq.(\ref{eq:buuuuu}) can be rewritten as
\begin{equation}
	\left[ \bar u_1 A P_L u_2 \right] \left[ \bar u_3 P_R B u_4 \right] = \frac{1}{2} [\bar u_1 A \gamma_\mu P_R B u_4] [\bar u_3 \gamma^\mu P_L u_2]\,,
	\label{eq:Fierz}
\end{equation}
which is the identity we wanted to prove. Notice that, unlike spinors,
fermionic fields anticommute, so their version of the Fierz Identity is
\begin{equation}
	\left[ \bar \psi_1 A P_L \psi_2 \right] \left[ \bar \psi_3 P_R B \psi_4 \right] = -\frac{1}{2} [\bar \psi_1 A \gamma_\mu P_R B \psi_4] [\bar \psi_3 \gamma^\mu P_L \psi_2]\,,
	\label{eq:Fierz_fields}
\end{equation}
with an extra minus sign.

\subsection{Fourier Transforms}

\subsubsection{Yukawa/Coulomb propagator}
\label{sec:FTCoulomb}

Let's compute the Fourier Transform of the Yukawa propagator,
\begin{equation}
	I \equiv \int \frac{\d^3 q}{(2\pi)^3}\, e^{i\vec q\, \vec r}\, \frac{1}{q^2 + \mu^2}\,,
\end{equation}
where $q \equiv |\vec q\,|$, which will also give the Fourier
Transform of the Coulomb propagator taking the limit $\mu\to 0$. In
spherical coordinates, $\d^3q = q^2\d q\, \d\cos\theta \, \d\phi$, we
get
\begin{equation}
	I = \frac{1}{(2\pi)^2} \, \int_0^\infty \d q\,\frac{q^2}{q^2+\mu^2} \int_{-1}^{1}\d\cos\theta e^{iqr\cos\theta}\,.
\end{equation}

Integrating over $\cos\theta$ we get
\begin{align}
	\nonumber I&= \frac{1}{(2\pi)^2} \int_0^\infty \d q\,\frac{q^2}{q^2+\mu^2}\, \frac{2\sin qr}{qr} = \frac{1}{(2\pi)^2} \text{ Im}\left\{\int_{-\infty}^\infty \d q\,\frac{q^2}{q^2+\mu^2}\, \frac{e^{iqr}}{qr}\right\} =\\
	&= \frac{1}{(2\pi)^2r} \text{ Im}\left\{\int_{-\infty}^\infty \d y\,\frac{y}{y^2+\mu^2r^2}\, e^{iy}\right\} \equiv \frac{1}{(2\pi)^2r} \text{ Im}\left\{\int_{-\infty}^\infty \d y\, f(y) \right\}\,.
\end{align}

\begin{figure}[t]
	\centering
	\begin{tikzpicture}[line width = 1.5pt, scale = 1.3]
		% Axes:
		\begin{scope}
			\clip (-2.75,-1) rectangle (3,3);
			\draw[fermion](-3, 0) -- (7.91 ,0);
			\draw[fermion](0, -1) -- (0, 6.27);
		\end{scope}
		\node at (3, 0) [right] {Re$\{y\}$};
		\node at (0,3) [above] {Im$\{y\}$};
		%Poles:
		\begin{scope}[shift={(0, 0.3)}]
			\draw[white, fill = white] (0,0) circle (0.07);
			\draw[fill = black] (0,0) circle (0.05);
				\node at (0, 0)[above right]{$i\mu r$};
		\end{scope}
		\begin{scope}[shift={(0, -0.3)}]
			\draw[white, fill = white] (0,0) circle (0.07);
			\draw[fill = black] (0,0) circle (0.05);
				\node at (0, 0)[right]{$-i\mu r$};
		\end{scope}
		%Integration path:
		\draw[fermion, thick] (2:2.55) arc (2:178:2.55);
			\node at (45:2.55)[above right]{$C_\infty$};
		\draw[fermion, thick] (178:2.55) -- (2:2.55);
	  \end{tikzpicture}
	  \caption{Integration path (in the complex plane of the $y$
            variable) used in the Residue Theorem for the integral in
            expression (\ref{eq:Coulint}).}
	  \label{fig:Coulint}
\end{figure}
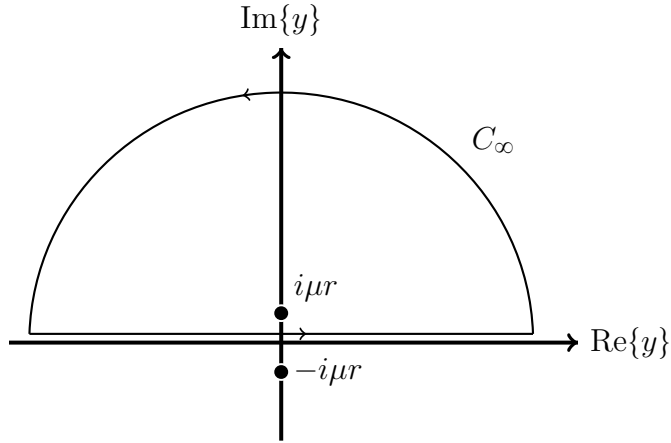

The integration of $f(y)$ along the circumference arc in
Fig.\ref{fig:Coulint} is zero, since it goes as $e^{-|y|}/|y|$ when
$y\to i\infty$. Therefore, using the Residue Theorem \cite{complexa},
\begin{equation}
	\int_{-\infty}^{\infty} \d y \, f(y) = \int_{C_\infty} \d y f(y) - 2\pi i\text{ Res}\left[ f(y), y= i \mu r \right] = -2\pi i \text{ Res}\left[ f(y), y= i \mu r \right]\,.
	\label{eq:Coulint}
\end{equation}

The pole of $f(y)$ in $y = i\mu r$ is simple, so we can compute the
residue as
\begin{equation}
	\text{ Res}\left[ f(y), y= i \mu r \right] = \lim_{y\to i\mu r} (y-i\mu r)f(y) =  \lim_{y\to i\mu r}  \frac{y}{(y + i\mu r)}\, e^{iy} = \frac{1}{2}\, e^{-\mu r}\,.
\end{equation}

Using this result, we can trivially write
\begin{equation}
	I  = \frac{1}{4\pi}\, \frac{e^{-\mu r}}{r}\,.
\end{equation}

\subsubsection{A spherical wave}
\label{sec:FTsphere}
Let's compute the integral
\begin{equation}
	I \equiv \frac{1}{4\pi}\int \d^3r\, e^{-i \vec q\, \vec r}\, \frac{e^{-\sqrt{t^\prime} r}}{r}\,,
\end{equation}
where the $4\pi$ factor has been introduced for convenience.

In spherical coordinates
$\d^3 r = r^2 \d r \, \d \Omega = r^2 \d r \,\d \cos \theta\, \d
\phi$,
\begin{equation}
	I = \frac{1}{2}\int_0^\infty \d r \, r\,e^{-\sqrt{t^\prime} r} \, \int_{-1}^1 \d\cos\theta \,e^{-i q r \cos\theta}\,,
\end{equation}
where the $\int \d \phi$ has been trivially computed and
$q\equiv \left| \vec q\, \right|$. The remaining angular integral is
also easy to calculate, so we can write
\begin{equation}
	I =\frac{1}{q} \int_0^\infty \d r \,e^{-\sqrt{t^\prime} r} \, \sin(qr)\,.
\end{equation}
%taking into account that $e^{iqr}-e^{-iqr} = 2i\sin(qr)$.
This integral can be computed integrating by parts:
\begin{align}
  \nonumber I &= \frac{1}{q} \int_0^\infty \d r \,\left\{ \frac{\d}{\d r}\left[ \frac{-1}{\sqrt{t^\prime}} e^{-\sqrt{t^\prime} r} \, \sin(qr)\right] - \left[ \frac{-1}{\sqrt{t^\prime}} e^{-\sqrt{t^\prime} r} \, \frac{\d}{\d r} \sin(qr)\right] \right\} =\\
  \nonumber &= 0 + \frac{1}{\sqrt{t^\prime}} \int_0^\infty \d r \,e^{-\sqrt{t^\prime} r} \, \cos(qr) = \\
  \nonumber &= \frac{1}{\sqrt{t^\prime}} \int_0^\infty \d r \,\left\{ \frac{\d}{\d r}\left[ \frac{-1}{\sqrt{t^\prime}} e^{-\sqrt{t^\prime} r} \, \cos(qr)\right] - \left[ \frac{-1}{\sqrt{t^\prime}} e^{-\sqrt{t^\prime} r} \, \frac{\d}{\d r} \cos(qr)\right] \right\} = \\
              &= \frac{1}{t^\prime} \left\{ 1 - q \int_0^\infty \d r \,e^{-\sqrt{t^\prime} r} \, \sin(qr) \right\}
\end{align}
Taking into account the fact that $q^2 \equiv \vec {q\,}^2 = -t$, this
last relation can be written as
\begin{equation}
    I = \frac{1}{t^\prime} (1+t I) \hspace{1cm} \longrightarrow \hspace{1cm} (t^\prime - t)I=1\,.
\end{equation}
Therefore, we have proved the relation
\begin{equation}
    \frac{1}{t^\prime - t} = \frac{1}{4\pi} \int \d^3r\, e^{-i \vec q\, \vec r}\, \frac{e^{-\sqrt{t^\prime} r}}{r}\,.
    \label{eq:FTSW}
\end{equation}

\subsection{Integrals for the absorptive part}
\label{sec:intsuch}

In this appendix we are going to calculate the integrals
\begin{subequations}
    \begin{align}
      &I \equiv  \int \d^4 k\, \delta (k^2) \delta(\bar k^2) = \frac{\pi}{2}\,, \\
      &I^\mu  \equiv\int \d^4 k\, \delta (k^2) \delta(\bar k^2)\,k^\mu = \frac{\pi}{4}\,q^\mu\,, \\
      &I^{\mu\nu} \equiv\int \d^4 k\, \delta (k^2)
        \delta(\bar k^2)\,k^\mu k^\nu =
        \frac{\pi}{6}\left( q^\mu q^\nu -
        \frac{1}{4}\, s\, g^{\mu\nu} \right)\,,
    \end{align}
\end{subequations}
where $\bar k = q-k$ and $q^2 = s$.

Let's consider the first one. Using $k^2 = E^2 - {\vec k\,}^2$ in the
first delta function, we compute the $E \equiv k^0$ integral,
\begin{equation}
    I = \int \frac{\d^3 k}{2E}\, \delta(\bar k^2) = \int \frac{\d^3 k}{2E}\, \delta\left[q^2 - 2(kq) \right]\,.
\end{equation}

We can evaluate the integral in the CM reference frame, where
$q^\mu = ( \sqrt{s}, \vec 0 \,)$, so that
\begin{equation}
    I = \frac{1}{2}\int \d \Omega\, \d E\, E\, \delta\left(s - 2E\sqrt{s} \right) = \frac{1}{2} \int \d\omega \left. \frac{E}{2\sqrt{s}} \right|_{E = \sqrt{s}/2} = \frac{1}{8}\int\d\Omega = \frac{\pi}{2}\,,
\end{equation}
as we wanted to prove.

Due to Lorentz covariance, the $I^\mu$ integral must be of the form
\begin{equation}
    I^\mu  \equiv \int \d^4 k\, \delta (k^2) \delta(\bar k^2)\,k^\mu= A \, q^\mu\,.
\end{equation}
Multiplying this relation by $q_\mu$, we get
\begin{equation}
    A\, q^2 = \int \d^4 k\, \delta (k^2) \delta(\bar k^2)\,(k\bar k) = \frac{1}{2}\,q^2\, I =\frac{\pi}{4}\,q^2\,,
\end{equation}
so
\begin{equation}
    I^\mu = \frac{\pi}{4}\, q^\mu\,.
\end{equation}

Finally, we can also use Lorentz covariance to write
\begin{equation}
    I^{\mu\nu}  \equiv \int \d^4 k\, \delta (k^2) \delta(\bar k^2)\,k^\mu k^\nu= A g^{\mu\nu}+ B q^\mu q^\nu\,.
\end{equation}

We need to multiply by $g_{\mu\nu}$ and $q_\mu q_\nu$ to determine $A$
and $B$,
\begin{subequations}
    \begin{align}
      &g_{\mu\nu}I^{\mu\nu} = 4 A + q^2 B = \int \d^4 k\, \delta (k^2) \delta(\bar k^2)\,k^2 = 0\,,\\
      &\frac{q_\mu q_\nu}{q^2} I^{\mu\nu} = A + q^2 B
        = \frac{1}{q^2} \int \d^4 k\, \delta (k^2)
        \delta(\bar k^2)\,(k\bar k)^2 = \frac{1}{4}\,
        ^2\, I = \frac{\pi}{8} q^2\,,
    \end{align}
\end{subequations}
so we just need to solve the algebraic system of equations
\begin{align}
  \nonumber 4 A + q^2 B &= 0\,,\\
  A + q^2 B &= \frac{\pi}{8}\, q^2\,,
\end{align}
which gives
\begin{equation}
    A = -\frac{\pi}{24}\, q^2\,, \hspace{4cm}	B = \frac{\pi}{6}\,.
\end{equation}

Therefore,
\begin{equation}
    I^{\mu\nu} = \frac{\pi}{6} \left( q^\mu q^\nu - \frac{1}{4}\, q^2\, g^{\mu\nu}  \right)\,.
\end{equation}

\subsection{Integral for the interaction potential}
\label{sec:AppI1}
We are interested in calculating a primitive $P(t)$ of
\begin{equation}
    \int \d t\, t \,  e^{-\sqrt{t} \,r}\,,
\end{equation}
which can be obtained quite straightforwardly using
$q \equiv \sqrt{t}$,
\begin{align}
  \nonumber P(t) &= \int \d t\, t \,  e^{-\sqrt{t} \,r} = 2 \int\,\d q\, q^3 \,  e^{- q r} = \\
  \nonumber &= -2\, \frac{\d^3}{\d r^3} \int \d q\, e^{-q r} = \\
  \nonumber &= 2 \, \frac{\d^3}{\d r^3} \left[ \frac{e^{-qr}}{r} \right] = \\
  \nonumber &= 2 \, \frac{\d^2}{\d r^2} \left[ \left( - \frac{1}{r^2} - \frac{q}{r} \right) e^{-qr} \right] = \\
  \nonumber &= 2 \, \frac{\d}{\d r} \left[ \left( \frac{2}{r^3} + \frac{2q}{r^2} + \frac{q^2}{r} \right) e^{-qr} \right] = \\
  \nonumber &=  -\left( \frac{12}{r^4} + \frac{12q}{r^3} + \frac{6q^2}{r^2} + \frac{2q^3}{r}\right) e^{-qr}  = \\
                 &= -\left( \frac{12}{r^4} + \frac{12\sqrt{t}}{r^3} + \frac{6t}{r^2} + \frac{2\sqrt{t^3}}{r}\right) e^{-\sqrt{t}\,r}\,.
\end{align}

%%%%%%%%%%%%%%%%%%%%%%%%%%%%%%%%%%%%%%%%%%%%%%
% FIGURAS Y TABLAS
%%%%%%%%%%%%%%%%%%%%%%%%%%%%%%%%%%%%%%%%%%%%%%
% \newpage
% \listoffigures
% \newpage
% \listoftables

%%%%%%%%%%%%%%%%%%%%%%%%%%%%%%%%%%%%%%%%%%%%%%
% ABREVIATURAS
%%%%%%%%%%%%%%%%%%%%%%%%%%%%%%%%%%%%%%%%%%%%%%
% \newpage
% \addcontentsline{toc}{part}{Abreviaturas}
%
% \section*{Abreviaturas}
%
%\begin{tabular}{l l}
%PDF		&Función de densidad de probabilidad
%\end{tabular}

%%%%%%%%%%%%%%%%%%%%%%%%%%%%%%%%%%%%%%%%%%%%%%
% BIBLIOGRAFÍA
%%%%%%%%%%%%%%%%%%%%%%%%%%%%%%%%%%%%%%%%%%%%%%

\end{document}